\documentclass[fleqn,10pt]{wlscirep}
\usepackage[utf8]{inputenc}
\usepackage[T1]{fontenc}
\usepackage{lineno}

\title{Cometary impacts an unlikely pathway towards the origins of life on Earth}

\author[1,*]{Richard~J.~Anslow}
\author[1]{Catriona~H.~McDonald}
\author[1]{Amy~Bonsor}
\author[2]{Paul~B.~Rimmer}
\author[3]{Auriol~S.~P.~Rae}
\affil[1]{Institute of Astronomy, University of Cambridge, Madingley Road, Cambridge, CB3 0HA, UK}
\affil[2]{Cavendish Laboratory, University of Cambridge, JJ Thomson Ave, Cambridge, CB3 0HE, UK}
\affil[2]{School of Geosciences/Edinburgh Centre for Planetary Sciences, University of Edinburgh, Edinburgh, EH9 3FE, UK}

\affil[*]{rja92@cam.ac.uk}

\begin{abstract}
An early period of cometary bombardment has long been speculated to provide an important source of prebiotic feedstock molecules, required for the origins of life on Earth. 
In support of this interpretation, comets boast a rich diversity of volatile organic molecules, with many thought to be limiting for prebiotic chemistry. Modelling studies suggest, however, that a significant proportion of these molecules will be destroyed during hypervelocity impact.
Here, we revisit the ability of comets to deliver prebiotic feedstock molecules in the high concentrations required for prebiotic chemistry, leveraging recent advances in cometary impact simulations, comet-atmosphere interactions, and dynamical modelling of short-period comets' origin and evolution. 
Taken together, the rapid decline in Earth's cometary impact bombardment and the efficient destruction of fragile cometary organics indicate that the successful delivery of prebiotic feedstock molecules was an exceedingly rare event.
Our results therefore challenge the long-held assumption that cometary impacts played an important role in the origins of life on Earth.
\end{abstract}

\begin{document}

\flushbottom
\maketitle

\thispagestyle{empty}

\section*{Introduction}

The origin of life on Earth presumably resulted from prebiotic chemistry, the synthesis of key molecular building blocks from an initial reservoir of reactive bioessential feedstock molecules\cite{Sasselov2020}. Hydrogen cyanide (HCN) has emerged as a key feedstock molecule, with its high-energy triple (${\rm C} \equiv {\rm N}$) bond able to thermodynamically drive subsequent chemistry, and assumes a central role in several prebiotic syntheses\cite{Powner2009,Patel2015,Becker2019}. Recent experimental studies demonstrate that high concentrations ($\sim$~10~mM) of HCN can synthesise many of the building blocks of life, including lipids, sugars, nucleobases, and nucleotides\cite{Powner2009,Patel2015,Xu2018,Becker2019,Foden2020}. Yet, the source of such a concentrated reservoir of feedstock molecules, such as HCN, on the early Earth is still debated.

Sources of HCN (and other C, H, N, and O molecules) have been historically divided into two categories: endogenous synthesis in the early terrestrial atmosphere, and exogenous delivery in the form of comets, asteroids, and interplanetary dust particles (IDPs)\cite{ChybaSagan1992}. The former, endogenous synthesis, is particularly sensitive to the geological context of prebiotic chemistry, with the yield of HCN production substantially higher in more reducing atmospheres\cite{Zahnle1986,RimmerRugheimer2019}. There is, however, substantial geological evidence against a reducing early atmosphere\cite{AulbachStagno2016,CatlingKasting2017}. Although geological processes such as serpentinization may help reduce an initially oxidised atmosphere\cite{Paschek2025},
even in the transiently reducing atmospheres afforded by the late accretion of large iron-rich planetesimals, rain-out yields steady-state HCN concentrations orders of magnitude lower than used in prebiotic experiment\cite{Zahnle2020,Wogan2023}. Such planet-scale events will deliver photochemically synthesised HCN to all exposed land on the early Earth, yet will require the subsequent stockpiling and concentration of cyanide (perhaps as ferrocyanide complexes) for subsequent utility in prebiotic chemistry\cite{Xu2018,Sasselov2020}.

In contrast, an exogenous source of Earth's prebiotic inventory -- a suggestion now more than a century old\cite{ChamberlinChamberlin1908,Oro1961} -- naturally bypasses any dependence on atmospheric oxidation state, and has the potential to deliver feedstock molecules in high concentration to local environments. Chyba \& Sagan\cite{ChybaSagan1992} demonstrated, using analytical fits to the lunar cratering record, that the exogenous delivery of carbon would be dominated by IDPs, with the cometary delivery of organic carbon insignificant on a global scale. The carbon and nitrogen content of IDPs is, however, dominated by insoluble complex aromatic hydrocarbons\cite{Clemett1993}, rendering IDPs a highly unlikely source of the reactive organic precursors (such as HCN) required for synthetic prebiotic chemistry. The same is not true for comets, which exhibit a rich diversity of volatile organic molecules, including most significantly HCN at the $\sim$~0.1~\% level relative to water\cite{MummaCharnley2011}. Comets therefore have the potential to deliver as much as $\sim$~30~mM HCN to local environments, a concentration of immediate relevance for prebiotic chemistry, and have accordingly been the subject of significant attention in origins of life research\cite{Clark1988,Chyba1990,PierazzoChyba1999,ToddOberg2020,McDonald2025,Zellner2020}.

Two key lines of evidence suggest that, despite not appearing in all prebiotic chemical networks, the intact delivery of HCN represents an end-member case for the cometary delivery of prebiotic feedstock molecules. 
First, HCN is the only feedstock molecule observed in Solar System comets in prebiotically relevant concentrations\cite{MummaCharnley2011}; other key feedstock molecules such as cyanoacetylene (${\rm HC_3N}$) and cyanogen (${\rm NCCN}$) are found to be orders of magnitude less abundant\cite{MummaCharnley2011,Hanni2021}. Second, the triple (nitrile) bond in HCN is expected to confer much greater durability to impact pyrolysis in comparison to other (more complex) prebiotic feedstock molecules\cite{McDonald2025}. If it transpires that comets were unable to deliver a prebiotically relevant concentration of HCN, we expect the same to be true for other prebiotic feedstock molecules.

To assess the relative importance of cometary impacts as a source of concentrated feedstock molecules, it is therefore crucial to accurately determine the ability of comets to deliver their HCN intact to the surface of the early Earth. Therein lies the crux of long-standing objections to cometary delivery, which relate to both the destruction of comets during atmospheric passage\cite{Chyba1990,ChybaSagan1992}, and the thermal destruction of fragile organic species during hypervelocity impact\cite{PierazzoChyba1999,ToddOberg2020}.
Recent modelling efforts demonstrate that only comets larger than $\sim100$~m will survive passage through the early Earth's atmosphere\cite{Anslow2025_atmos}, and that the survival of HCN during hypervelocity impact has been historically overestimated\cite{McDonald2025}.
It is thus prudent to quantitively re-examine the cometary delivery of prebiotic feedstock molecules, given that multiple lines of evidence suggest its efficacy may have been historically overestimated.

The present study uses the results of N-body simulation of short-period comet formation and evolution\cite{Nesvorny2023,Nesvorny2017:SPCs} to reassess the potential role of cometary delivery in the origins of life on Earth. 
We leverage recent methodological advances in the modelling of cometary atmospheric entry\cite{Anslow2025_atmos} and the destruction of HCN during hypervelocity impact\cite{McDonald2025} to quantify the number of comets able to deliver key bioessential feedstock molecules in prebiotically relevant concentration. We find that constraints on comet size, velocity, and angle -- necessary for the intact survival of comets and their HCN during hypervelocity impact -- preclude a robust cometary source of concentrated prebiotic feedstock molecules. 
Thus, despite the rich diversity in volatile organic species observed in Solar System comets, our results challenge the hypothesis that comets played an important role in the origins of life on Earth.

\section*{Results}

\subsection*{Earth's early cometary bombardment}

Here, we use Earth's cometary bombardment chronology\cite{Nesvorny2023} as derived from the Nesvorn\'y et al.\cite{Nesvorny2017:SPCs} model of short-period comet formation and evolution. 
This model, following the long-term evolution of one million primordial cometesimals originating in a disk exterior to Neptune, is calibrated against several key observational constraints from the outer Solar System\cite{Nesvorny2017:SPCs,Nesvorny2019,Nesvorny2023}. This approach diverges from previous modelling efforts, which instead use the lunar crater record as an indirect proxy for Earth's early cometary bombardment [e.g.,\cite{Chyba1990,ChybaSagan1992,PierazzoChyba1999}]. This represents an important methodological distinction, given that it is now clear that the lunar crater record neither unambiguously constrains impactor composition\cite{Bailey1991}, nor provides a reliable record of the Moon's bombardment history pre-3.9~Gya\cite{Zellner2017}. Indeed, since the lunar crater record is consistent with an asteroidal origin\cite{Strom2005}, it appears unlikely that it will accurately constrain Earth's early cometary bombardment.
We note that we focus specifically on short-period comets, whose characteristic impact velocities are substantially lower than those of long-period comets\cite{HughesWilliams2000}, given that the survival of prebiotic feedstock molecules depends sensitively on impact velocity\cite{PierazzoChyba1999,ToddOberg2020,McDonald2025} (see\cite{methods} for further discussion).

To reproduce the observed distribution of modern short-period comets, such dynamical models require some mechanism to limit the physical lifetime of small comets as they are scattered into the inner Solar System\cite{DiSisto2009,Nesvorny2017:SPCs}.
This requirement is consistent with modern Jupiter-family comets (JFCs), which are observed to frequently undergo spontaneous fragmentation\cite{Sekanina1984,Boehnhardt2002,Jewitt2016}, and have physical lifetimes an order of magnitude shorter than their dynamical lifetimes\cite{LevisonDuncan1997}.

\begin{figure}[!t]
    \centering
    \includegraphics[width=0.6\textwidth]{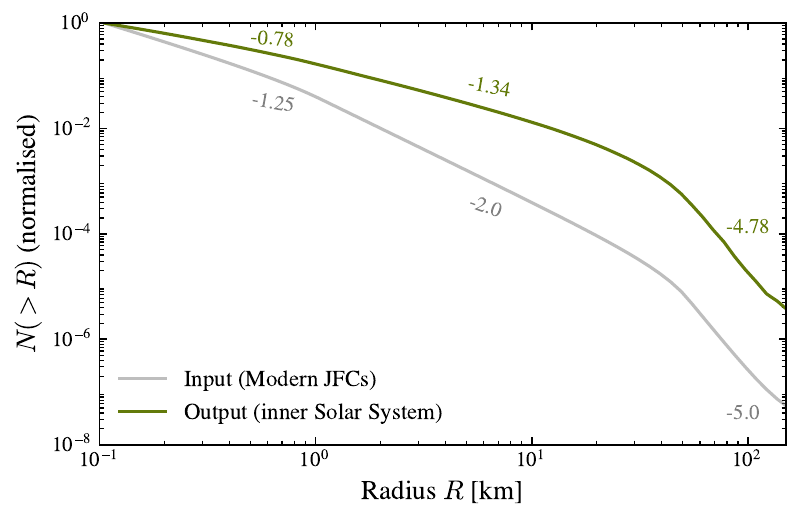}
    \caption{\textbf{Cumulative size-frequency distribution (SFD) of comets entering the inner Solar System subject to disruption via rotational instability.}  
    The slopes of the input (gray) and output (green) SFDs are annotated next to the curves for the size ranges $0.1-1$~km, $1-50$~km, and $50-150$~km. The strong size-dependence of rotational spin-up\cite{Jewitt2021} drives the preferential disruption of small comets, causing a significant decrease in the power-law slope at small radii.}
    \label{fig:SFD_evol}
\end{figure}

Numerical models were used to determine the effects of rotational instability on the size-frequency distribution (SFD) of comets scattered into the inner Solar System\cite{methods} .
The preferential disruption of small comets, due to the strong size-dependence of anisotropic outgassing torques, is found to drive order of magnitude reduction in the number of comets accreted by the early Earth (which we verify in the Supplementary Text is robust to this specific fragmentation model). This significant decrease in Earth's overall cometary impact flux is reflected in the substantial decrease in the slope of short-period comets' SFD at small radii (Fig.~\ref{fig:SFD_evol}).

\subsection*{The atmospheric destruction of small and oblique comets}

The low density (i.e., high porosity) and weak mechanical strength of comets\cite{Groussin2019} combine such that comets are particularly vulnerable to destruction during atmospheric passage -- a consequence of the rapid, and violent transfer of comets' momentum to the surrounding atmosphere\cite{HillsGoda1993,Anslow2025_atmos}. This is expected to be most significant for small and oblique comets, which must survive passage through comparatively massive atmospheric columns, leading to the destruction of even km-scale comets\cite{Anslow2025_atmos}. Earth’s atmosphere therefore acts as a filter -- preferentially destroying small and oblique comets -- and thus, biasing the population of comets able to deliver prebiotic feedstock molecules to local surficial environments.

To determine the effects of atmospheric entry on the cometary impact flux reaching Earth's surface, we use the \textsc{atmosentry} code\cite{Anslow2025_atmos} to model the deceleration and fragmentation of comets during passage through Earth's atmosphere. As discussed in Anslow et al.\cite{Anslow2025_atmos}, a key environmental parameter controlling the survival of comets is atmospheric surface pressure. Geological constraints on the Hadean atmosphere are, however, scarce and remain subject to debate\cite{CatlingZahnle2020}. We therefore adopt a 1~bar atmosphere, which we expect to present a best-case scenario for cometary survival\cite{methods}.

\begin{figure}[!t]
    \centering
    \includegraphics[width=\textwidth]{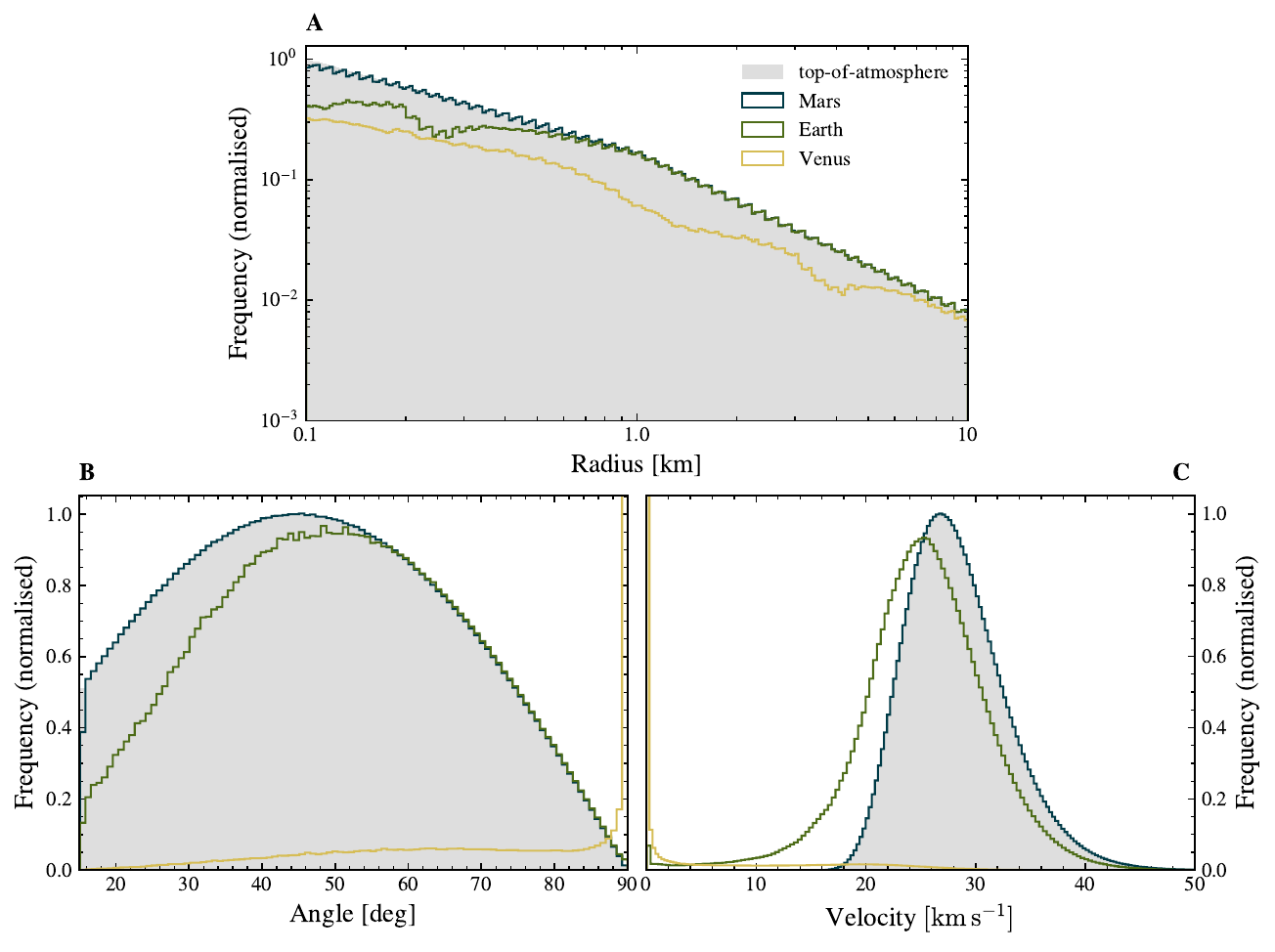}
    \caption{\textbf{The influence of atmospheric entry on short-period comets' size, impact angle and impact velocity distributions.} The gray shaded histograms show the initial size (\textbf{A}), angle (\textbf{B}), and velocity (\textbf{C}) distributions of comets incident at the top of Earth's atmosphere. The corresponding distributions following atmospheric entry are shown for modern-day terrestrial (green), Martian (blue) and Venusian (yellow) atmospheres.
    }
    \label{fig:TOA_surf_dists_terrestrial}
\end{figure}

As anticipated, the effects of atmospheric entry are most pronounced, on a population level, for small comets with initially oblique entry angles (Fig.~\ref{fig:TOA_surf_dists_terrestrial}). Comets smaller than $\sim100~$m, regardless of impact angle, are found to catastrophically fragment in explosive airbursts at high altitude, distributing their feedstock molecules over a large surface area. This represents a critical challenge for the successful delivery of prebiotic feedstock molecules, given that atmospheric entry biases the surviving population of comets against oblique impacts -- the small subset of comets able to deliver their prebiotic feedstock molecules, in high concentration, to Earth's surface\cite{PierazzoChyba1999,McDonald2025}.
Conversely, a by-product of atmospheric aerobraking is the production of a low velocity tail in the comets' impact velocity distribution, which is found to extend significantly below Earth's escape velocity. This, in contrast, will promote the intact delivery of fragile cometary organics\cite{PierazzoChyba1999,McDonald2025}.
Atmospheric entry thus not only limits the total number, but also significantly biases the distribution of comets reaching Earth's surface intact, with significant implications for the subsequent survival of prebiotic feedstock molecules during hypervelocity impact.

\subsection*{The inefficient delivery of prebiotic feedstock molecules to the early Earth}

To assess the ability of comets to deliver, intact, prebiotic feedstock molecules to Earth's surface, we employ Monte Carlo simulation, leveraging the recent advances in short-period comet formation and evolution, and modelling of cometary atmospheric entry described thus far.

We discretise Earth's cometary bombardment into 1~Myr time bins, generating the population of comets incident at the top-of-atmosphere from a Poisson distribution with mean number anchored to the Nesvorn\'y et al.\cite{Nesvorny2023} impact chronology [see\cite{methods} for more details]. 
The population of comets (and fragments thereof) reaching Earth's surface intact, surviving atmospheric entry, is determined using the numerical model \textsc{atmosentry} as described above. For each comet reaching the surface, the amount of HCN surviving the high-temperature conditions associated with hypervelocity impact is finally calculated following McDonald et al.\cite{McDonald2025}.

Our simulations record the total mass accretion of $(2.63\pm0.19)\times10^{18}$~kg, which is consistent with constraints on total cometary accretion from noble gas isotopic ratios [$\sim10^{18}$~kg;\cite{Marty2017,Cassata2025}], lending independent support to our parametrised model of Earth's cometary bombardment.
Yet, stringent constraints on comet size and entry angle -- required for the survival of comets during atmospheric entry -- are largely in tension with the survival of HCN during cometary impact\cite{Anslow2025_atmos,McDonald2025}. The successful delivery of HCN is accordingly limited to a very narrow region of parameter space (Supplementary Text), which as we show below, renders comets an insignificant source of concentrated HCN on both global and local scales.

On a global scale we find, in agreement with the results of Todd \& {\" O}berg\cite{ToddOberg2020}, that comets are unable to deliver a prebiotically relevant concentration of HCN to the early Earth. Cumulative, intact, HCN delivery is limited to only $2.65\times10^{13}$~kg. Diluted in 1 terrestrial ocean of water --  totally neglecting the subsequent hydrolysis and degradation of HCN -- this corresponds to a maximum concentration of $\sim$~0.75~${\rm \mu M}$, which is orders of magnitude below that used in prebiotic experiment. Alternatively, assuming a 1~bar ${\rm N}_2$-dominated atmosphere, this corresponds to a maximum HCN mixing ratio of order $10^{-6}$ -- several orders of magnitude below that possible via photochemistry in both neutral\cite{RimmerRugheimer2019} and transiently reducing\cite{Zahnle2020,Wogan2023} atmospheres.

The potential role of comets in the origins of life on Earth is therefore restricted to the delivery of HCN, in high concentration, to local subaerial environments. During our Monte Carlo simulations of Earth's cometary bombardment, we thus record the maximum concentration of HCN delivered per cometary impact -- under the assumption of dilution in only the comet's water ice. 
We assume a minimum HCN concentration of $10~$mM is required for subsequent prebiotic chemistry, based on recent laboratory experiment [e.g,.\cite{Xu2018}], however note our conclusions are robust to this choice (Supplementary Text).
In any case, we expect this to represent a significant upper limit on actual HCN concentration -- and thus best-case scenario for subsequent prebiotic chemistry -- given that several environmental processes will act, quickly, to dilute cometary HCN. In addition to precipitation, the thermobaric phase of impact cratering will unavoidably disrupt the local hydrosphere (or cryosphere)\cite{Osinski2020}. A salient outcome of high-velocity cometary impacts, irrespective of the delivery of HCN, is thus the rapid influx of local groundwater into the newly-formed crater-lake, rapidly diluting surviving cometary HCN below prebiotically relevant concentrations.

To obtain meaningful results, given uncertainty in the extent of subaerial land on the early Earth\cite{Korenaga2021}, we adopt the McCulloch \& Bennett\cite{McCullochBennett1993} model of continental growth as a proxy for exposed landmass (following\cite{Pearce2017}), noting that our results are insensitive to this choice (Supplementary Text). This is likely a significant overestimate of exposed landmass on the Hadean Earth, given the non-monotonic history of ocean volume\cite{BadaKorenaga2018}.
Indeed, it is plausible that all continental crust will have been submarine during the early Hadean -- during which period Earth's cometary bombardment was most intense -- limiting exposed landmass to hotspot oceanic islands\cite{Korenaga2017,Korenaga2021}. This is not a problem for the purposes of this work: by adopting an optimistic model of continental growth, and analogously by recording the maximum HCN concentration post-impact, we ensure our results constitute a best-case scenario for cometary delivery.

\begin{figure}[!t]
    \centering
    \includegraphics[width=0.6\textwidth]{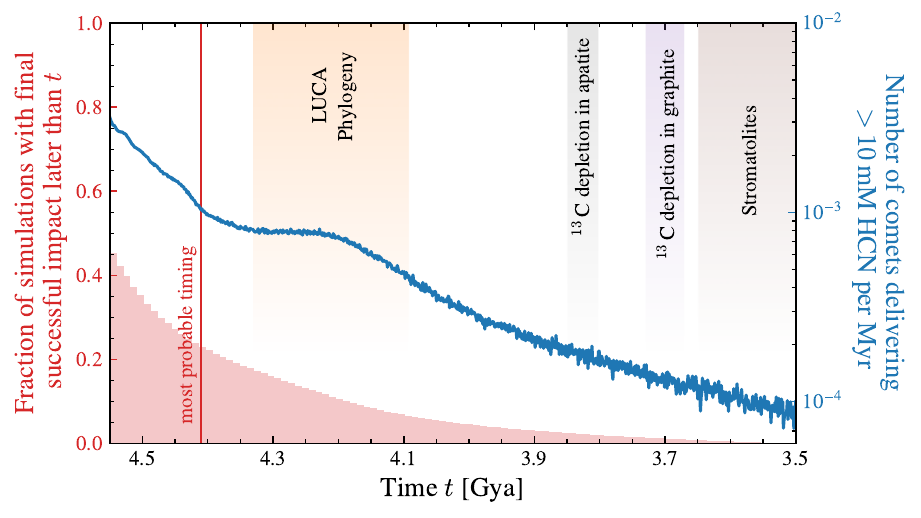}
    \caption{\textbf{Results from 100,000 Monte Carlo realisations of Earth's cometary bombardment history.}
    The number of comets delivering a prebiotically relevant concentration ($\geq10$~mM) of HCN per 1~Myr time-bin, following the Nesvorn\'y et al.\cite{Nesvorny2023} impact chronology, is shown in blue.
    We consider only cometary impacts onto exposed subaerial land and use the McCulloch \& Bennett\cite{McCullochBennett1993} model of continental growth [see\cite{methods}], and assume optimistically that surviving HCN is diluted in only the comet's water ice.
    The median number of successful cometary impacts per Monte Carlo realisation is $0$, with $54~$\% of our simulations recording no successful cometary impact.
    Successful cometary impacts are overwhelmingly likely to occur early in Earth's history (shown in red), with less than $5~$\% of our simulations recording the delivery of $>10$~mM HCN later than $4.03$~Gya. The most probable timing of successful cometary impact is $4.41^{+0.11}_{-0.29}$~Gya -- more than 100~Myr earlier than the first, indirect evidence of life on Earth at $\sim4.2~{\rm Gya}$ ($4.09-4.33~{\rm Gya}$; orange shaded region)\cite{Moody2024}. Error bars report the 16th–84th percentile range of the last successful impact, for simulations recording at least one successful cometary impact.
    }
    \label{fig:successful_impacts}
\end{figure}

Despite representing a best-case scenario for cometary delivery, our Monte Carlo simulations of Earth's early cometary bombardment record, on average, zero cometary impacts delivering a prebiotically relevant concentration of HCN ($>10$~mM). This reflects significant inefficiency in the delivery of HCN, given that Earth accreted more than $10^5$ ($D>100~$m) comets during this early period of intense impact bombardment. This inefficiency is driven primarily by the high characteristic impact velocity of short-period comets\cite{Nesvorny2023}, which is found to restrict the successful delivery of HCN to only highly-oblique impacts at the low-velocity tail of the comets' impact velocity distribution (Supplementary Text). This corresponds to a very small region of parameter space, which is only sampled at very early times when the cometary impact flux is highest (Fig.~\ref{fig:successful_impacts}) -- a conclusion we find to be robust to the assumed dynamical model of Earth's cometary bombardment (Supplementary Text). 
Consequently, the most probable timing of successful cometary impact is in the early Hadean, $4.41^{+0.11}_{-0.29}$~Gya, more than 100~Myr earlier than the estimated emergence of the last universal common ancestor (LUCA) at $\sim$~4.2~Gya\cite{Moody2024}.

The inability of comets to robustly deliver HCN in high concentration is further compounded by the coincidence of Earth's most intense cometary bombardment with the limited availability of exposed subaerial land during the early Hadean\cite{BadaKorenaga2018,Korenaga2021}. Owing to the near absence of continental crust during the early Hadean\cite{Korenaga2021}, our simulations record a significant number of low-velocity cometary impacts onto Earth's early oceans. Any surviving HCN from these impacts will have been quickly diluted and lost to hydrolysis, unable to trigger subsequent prebiotic chemistry. Taken together, our results demonstrate that restrictions on comet size, velocity, and angle -- imposed by constraints on the survival of comets and their HCN during impact with Earth -- combine such that the successful delivery of prebiotic feedstock molecules was an exceedingly rare event, with the majority of our Monte Carlo simulations -- even in this best-case scenario -- recording zero successful cometary impacts.

\subsection*{Comparative analysis of inner Solar System cometary bombardment}

Despite lacking any evidence of life, the dynamical simulations of Nesvorn\'y et al.\cite{Nesvorny2023} allow us to also consider the prebiotic implications of this early period of cometary bombardment for Mars and Venus, our closest planetary neighbours.
Given sparse paleoenvironmental constraints for both Mars and Venus, we adopt modern-day environmental parameters to enable direct comparison with Earth. Critically, despite abundant geological evidence of fluvial processes on Mars\cite{Carr1987,McLennan2019}, the nature and evolution of surficial water on Mars remain subject to debate, while more fundamental uncertainty remains regarding the existence of liquid-water oceans on Venus\cite{Way2016,Turbet2021,Constantinou2025:DryVenus}. Accordingly, we make no assumption regarding the extent of subaerial landmass, and focus only on the cometary flux surviving atmospheric entry, and HCN surviving hypervelocity impact.

\begin{figure}[!t]
    \centering
    \includegraphics[width=0.6\textwidth]{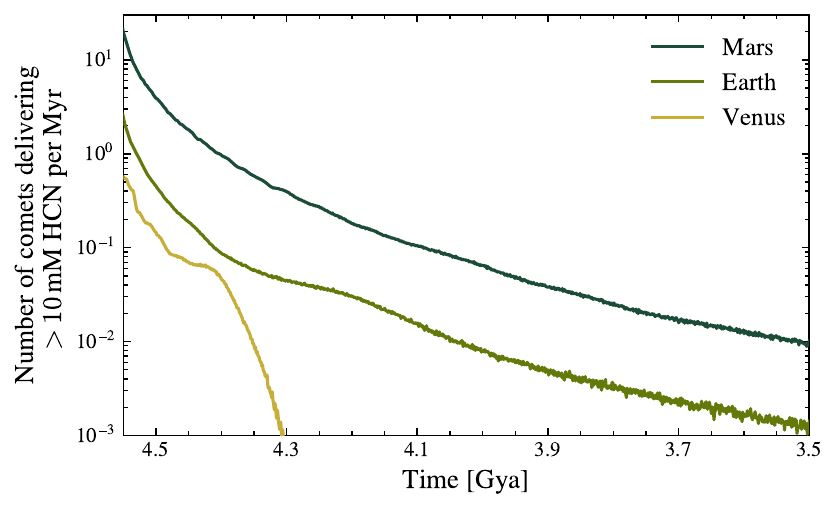}
    \caption{
    \textbf{Results from Monte Carlo simulations of the early cometary bombardment of Mars (blue), Earth (green) and Venus (yellow).}
    The number of cometary impacts delivering a prebiotically relevant concentration ($>10$~mM) of HCN as a function of time, in bins of 1~Myr, averaged over $10^5$ Monte Carlo realisations of the Nesvorn\'y\cite{Nesvorny2023} impact chronology.
    These results represent an upper limit on the number of environments able to support subsequent prebiotic chemistry, making no assumption about the availability of subaerial landmass. 
    }
    \label{fig:num_impacts_terrestrial_comparison}
\end{figure}

We find that cometary delivery is far more effective onto Mars (Fig.~\ref{fig:num_impacts_terrestrial_comparison}). Despite an overall impact flux $\sim5\times$ lower than on Earth and Venus\cite{Nesvorny2023}, Mars receives $744\pm27$ impacts delivering $>10\,$mM HCN, compared with $87\pm9$ for Earth (ignoring the extent of subaerial landmass) and $24\pm5$ for Venus. This efficiency is driven primarily by lower cometary impact velocities -- a consequence of Mars' lower escape velocity and larger semi-major axis\cite{Anslow2023:DeliveryExoplanets} -- and Mars' tenuous atmosphere. These combine such that almost all $D>100~$m comets are able to survive atmospheric entry (Fig.~\ref{fig:TOA_surf_dists_terrestrial}), with a much larger fraction of these comets delivering their HCN intact.
Even if Mars' early atmospheric surface pressure approached 1~bar, to facilitate stable surface liquid water\cite{Wordsworth2016:EarlyMarsClimate}, the number of successful impacts ($448\pm21$) remains more than five times greater than on Earth. In stark contrast, Venus' dense atmosphere destroys the majority of incident comets -- even those several kilometres in diameter -- making the successful delivery of HCN substantially more challenging than onto Earth.

This simple comparison between the Solar System's terrestrial planets suggests that, all else being equal, Mars would have been the more favourable target for the cometary delivery of prebiotic feedstock molecules, and if cometary delivery is an important prerequisite step for prebiotic chemistry, strengthens the possibility that life could have emerged on Mars. 
We caution, however, that this analysis is agnostic to global environmental conditions, including most importantly the availability of subaerial land, and thus reiterate that these results represent a best case scenario for subsequent prebiotic chemistry. Indeed, the dearth of exposed land on the Hadean Earth decreases the median number of successful impacts from 87 to 0. Clearly, a similar reduction might also be realised in the case of a warm and wet (pre-)Noachian Mars, harbouring a globally significant reservoir of surficial water.

\section*{Discussion}

\subsection*{Implications for the origins of life on Earth}

Although Solar System comets boast a rich diversity of volatile organic species -- including crucially many prebiotic feedstock molecules in high concentration\cite{MummaCharnley2011} -- we have shown via Monte Carlo simulation of Earth's cometary bombardment that these comets were, on the whole, unable to deliver these fragile organic molecules to Earth's surface intact. Our results demonstrate that, on both global and local scales, cometary impacts provided an insignificant source of the concentrated feedstock molecules required for prebiotic chemistry, challenging the long-held hypothesis that comets played an important role in the origins of life on Earth\cite{ChamberlinChamberlin1908,Oro1961,Clark1988,Chyba1990,PierazzoChyba1999}.

On a global scale, the intact delivery of HCN is orders of magnitude below what can be produced endogenously in even the most unfavourable (i.e., oxidising) atmospheric conditions. On a local scale, we find that in the best-case scenario in which surviving HCN is diluted within only the comet's water ice, the median number of impacts delivering a prebiotically relevant concentration of HCN is zero (Fig.~\ref{fig:successful_impacts}). Among the 46~\% of simulations that record at least one successful impact, the most probable timing of the last successful impact is $4.41^{+0.11}_{-0.29}$~Gya. These results reflect the rapid decline in Earth's cometary bombardment\cite{Nesvorny2023}, the limited availability of subaerial land\cite{Korenaga2021}, and the exceedingly small region of parameter space in which HCN can survive hypervelocity impact\cite{McDonald2025}.

Multiple lines of evidence suggest that these results represent a best-case scenario for the cometary delivery of HCN, and indeed other prebiotic feedstock molecules.
Most significantly, whenever faced with loosely constrained environmental parameters throughout our analysis -- such as atmospheric surface pressure or the extent of subaerial land on the Hadean Earth -- we have adopted values (within plausible limits) to maximise the number of successful cometary impacts. 
Moreover, it is likely that we significantly overestimate HCN survival during cometary impacts. We adopt a parametrised model of HCN survival directly from McDonald et al.\cite{McDonald2025}, which is based on idealised simulations that include only one destruction pathway for HCN, and crucially, neglect comets' internal strength and porosity -- both of which are expected to increase peak temperatures during hypervelocity impact. While the destruction of HCN will produce a variety of reactive intermediates -- with the impact-driven synthesis of nucleobases and sugars demonstrated in highly reducing atmospheres\cite{Ferus2020} -- the primary decomposition products of HCN in the presence of water are CO and CO$_2$\cite{Knizek2024}, which are not of particular prebiotic significance.
We expect this picture to be significantly worse for more complex prebiotic feedstock molecules: HCN is not only the only feedstock molecule found in comets at prebiotically relevant concentrations \cite{MummaCharnley2011}, but is, by virtue of its nitrile bond, expected to be significantly more resistant to thermal destruction during impact than other feedstock molecules \cite{McDonald2025}.

It is therefore overwhelmingly likely, based on the results of our Monte Carlo simulations, that in practice no comets delivered a prebiotically relevant concentration of HCN, in tension with the long-held hypothesis that cometary impacts may have played an important role in the origins of life on Earth\cite{ChamberlinChamberlin1908,Oro1961,Clark1988,Chyba1990,PierazzoChyba1999}. Nonetheless, we cannot discount the possibility -- in the spirit of Clark\cite{Clark1988} -- that during the early Hadean, the fortuitous collision of a low-velocity, highly oblique comet may have delivered almost all of its HCN intact. It is therefore critical to address the additional environmental barriers that subsequent prebiotic chemistry would face in the event of such a stochastic, successful cometary impact.

The most immediate and decisive threat to subsequent prebiotic chemistry is the environmental dilution of cometary HCN.
During hypervelocity impact, a comet's kinetic energy is rapidly transferred to the Earth, with peak pressures of order $10-100$~GPa driving the shock metamorphism (fracture, deformation and melting) of target rocks\cite{Melosh1989:ImpactCratering}. A significant outcome of this initial and violent `thermobaric' phase of impact cratering is the disruption of the local hydrosphere (or cryosphere)\cite{Osinski2020}. While the subsequent evolution of post-impact crater environments is enormously diverse -- sensitive to crater size, target lithology, and local environment\cite{Osinski2020} -- the influx of groundwater (if present), coupled with routine precipitation and surficial runoff, will act to rapidly dilute cometary HCN on a timescale of only days to weeks. Further compounded by the subsequent hydrolysis of HCN into formamide\cite{ToddOberg2020}, even an initially prebiotically competitive concentration of HCN will be quickly driven below functional levels.

Subsequent prebiotic chemistry must therefore succeed on short timescale, while cometary HCN remains at a prebiotically functional concentration. We now demonstrate, via a heuristic toy model, that this places strict -- and, we argue, implausible -- constraints on the efficiency of prebiotic chemistry, challenging the long-held suggestion that comets played an important role in the origins of life on Earth.

To place this on a more quantitative footing, we make the simplifying assumption that a successful cometary impact is a necessary prerequisite step for the origins of life. Given that a single cometary impact will never deliver the full complement of bioessential feedstock molecules required for any prebiotic chemical network, we further assume there is some (unknown) probability of abiogenesis per unit time, $p_{\rm abio}$. If prebiotic chemistry can only occur within $\tau_{\rm HCN}$ of successful cometary impact -- i.e., the time window during which HCN remains in sufficiently high concentration -- then the probability this chemistry succeeds following a single cometary impact is thus
\begin{equation}
    p_{\rm suc} = 1 - \exp{\left(-p_{\rm abio}\tau_{\rm HCN}\right)} \simeq p_{\rm abio}\tau_{\rm HCN},
\end{equation}
assuming $p_{\rm abio}\tau_{\rm HCN} \ll 1$.

Given that our Monte Carlo simulations reveal that the successful cometary delivery of HCN was an exceedingly rare event, we are able to estimate the likelihood of prebiotic chemistry succeeding on the early Earth. Specifically, we derive a heuristic origins of life timescale, $\tau_{\rm ool}$, representing the expected time required for prebiotic chemistry to succeed given both the rarity of successful cometary impact, and the rapid environmental dilution of cometary HCN. If this timescale is significantly longer than the duration of Earth's intense cometary bombardment, this would quantitatively encode how unlikely it is that cometary impacts played an important role in the origins of life on Earth.

\begin{figure}[!t]
    \centering
    \includegraphics[width=0.6\textwidth]{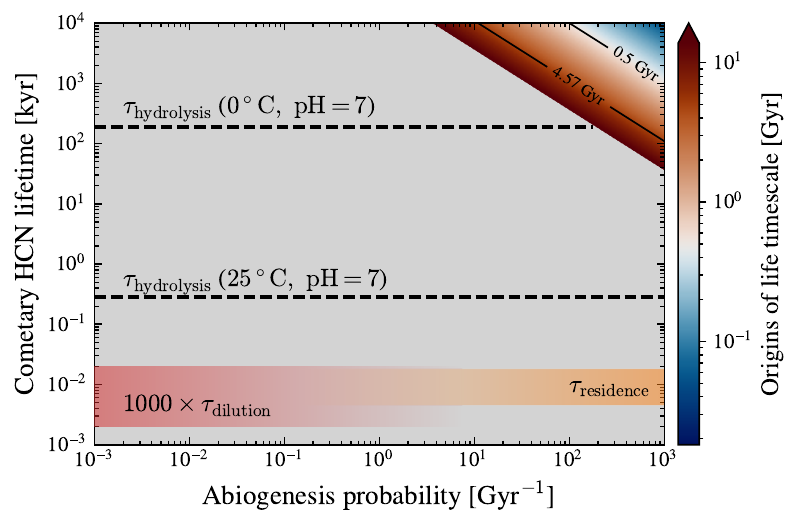}
    \caption{\textbf{The origins-of-life timescale is in excess of the age of the Solar System for plausible values of abiogenesis probability and cometary HCN lifetime.}
    A heuristic origins-of-life timescale, $\tau_{\rm ool}$, is calculated under the assumption that a successful cometary impact is a necessary prerequisite for the origins of life. Bayesian analyses, conditioned on the early emergence of life on Earth, support an abiogenesis probability in the range $10^{-3}<p_{\rm abio}<10^3~{\rm Gyr}^{-1}$\cite{SpiegelTurner2012,ChenKipping2018}. The lifetime of cometary HCN is expected to be determined by either the timescale of environmental dilution [\cite{Osinski2020}, red horizontal band], or the residence time of the crater-lake [\cite{Messager2016}, orange horizontal band]. The hydrolysis lifetime of HCN in aqueous solution at $0^\circ$C and $25^\circ$C [\cite{Miyakawa2002}, black dashed lines] represent significant upper limits. In all cases, this heuristic origins-of-life timescale is orders of magnitude larger than the duration of Earth's intense cometary bombardment: prebiotic chemistry was exceedingly unlikely to succeed following successful cometary impact.
    }
    \label{fig:ool_timescale_contours}
\end{figure}

The results of our Monte Carlo simulations demonstrate -- if they occurred at all -- that successful cometary impacts are overwhelmingly likely to have occurred early in Earth’s history, with less than $5\,$\% of simulations recording the successful delivery of HCN later than $4.03~$Gya (Fig.~\ref{fig:successful_impacts}). To first order, the number of successful impacts can be modelled as Bernoulli distributed with constant rate $r_{\rm imp} = T_{\rm imp}^{-1} \sim 2\times10^{-3}~{\rm Myr}^{-1}$, where $T_{\rm imp}\simeq 500~{\rm Myr}$ is the approximate duration of Earth's intense cometary bombardment. This heuristic origins of life timescale is thus given by
\begin{equation}
    \label{eq:origins_of_life_timescale}
    \tau_{\rm ool} \simeq \lambda_{\rm suc}^{-1} \simeq \dfrac{T_{\rm imp}}{p_{\rm abio} \tau_{\rm HCN}},
\end{equation}
which is shown in Fig.~\ref{fig:ool_timescale_contours} across a wide range of $p_{\rm abio}-\tau_{\rm HCN}$ parameter space. In the following paragraphs, we constrain plausible bounds for $\tau_{\rm HCN}$ and $p_{\rm abio}$, thereby allowing this illustrative calculation to quantitatively assess the potential role of cometary impacts in the origins of life on Earth.

As already discussed, there are several competing environmental and chemical processes that, together, will control the productive lifetime of a cometary crater-lake, $\tau_{\rm HCN}$. Of these, we expect the environmental dilution of cometary HCN to play the most decisive role, operating on a timescale of days to weeks following cometary impact\cite{Osinski2020}. If, however, supportive environmental conditions prevent the rapid dilution of HCN, the residence time of the crater-lake -- the average time water remains in the lake -- would instead be the more appropriate timescale to consider. Through analogy with modern lakes, whose residence times span several orders of magnitude, we expect this timescale to be of the order $1-10$~yr\cite{Messager2016}. Thus, while the hydrolysis of HCN into formamide will occur over much longer timescales $(\gtrsim{\rm kyr})$, dependent on both temperature and pH\cite{Miyakawa2002}, it is highly unlikely that appreciable cometary HCN will survive long enough for this to represent a dominant loss process.

Under the assumption that a successful cometary impact is a necessary prerequisite step for the origins of life, we identify the abiogenesis rate adopted and discussed in\cite{SpiegelTurner2012,ChenKipping2018} as a plausible, order-of-magnitude estimate for $p_{\rm abio}$. These Bayesian analyses, conditioned upon the early emergence of life on Earth, support an abiogenesis rate in the approximate range $10^{-3}<p_{\rm abio}<10^3~{\rm Gyr}^{-1}$\cite{SpiegelTurner2012,ChenKipping2018}. This would imply an origins of life timescale in excess of the age of the Universe for the plausible values of $\tau_{\rm HCN}$ considered above (Fig.~\ref{fig:ool_timescale_contours}). It is clear that we are fundamentally limited by our ignorance regarding most inputs required to determine such a quantity; this is particularly true of $p_{\rm abio}$, where we draw attention to the significant transfer of scale from planet-scale Bayesian analysis to local crater-lake environments. Nonetheless, reliance on a single cometary impact, and the short time-window available for prebiotic chemistry, combine such that the results of this heuristic exercise are robust to significant uncertainty in the abiogenesis rate, $p_{\rm abio}$.

It is therefore apparent, through Monte Carlo simulation of Earth's cometary bombardment, that synthetic prebiotic chemistry was exceedingly unlikely to succeed following cometary impact. While cometary HCN could, in principle, be subsequently concentrated and stored in ferrocyanide salts\cite{Sasselov2020}, it is non-trivial to subsequently release this cyanide in a reactive form\cite{Anslow2025_overlap}. This would likely require the thermal transformation of ferrocyanide salts via volcanism or lightning, introducing an additional reliance on supportive local environmental conditions to what is already a quantitatively implausible scenario. Moreover, this would remove the principal advantage attributed to cometary delivery -- the immediate delivery of a prebiotically relevant concentration of key feedstock molecules -- and so confers little benefit relative to an endogenous source of these molecules.  Despite not reaching concentrations immediately comparable to that used in prebiotic experiment\cite{Benner2020,Zahnle2020,Wogan2023}, an endogenous source would by contrast deliver these key feedstock molecules, in quasi steady-state, to all exposed subaerial land.

To summarise, our results demonstrate that Earth's cometary bombardment was unable to provide the concentrated reservoir of bioessential feedstock molecules required for synthetic prebiotic chemistry, challenging the long-held suggestion that cometary impacts may plausibly have played an important role in the origins of life on Earth. While we cannot formally exclude the possibility that a single fortuitous cometary impact delivered a prebiotically plausible concentration of HCN, we have demonstrated this would impose strict -- and, we argue, implausible -- constraints on the efficiency of subsequent prebiotic chemistry. Prebiotic chemistry would be required to succeed on short timescales, before the rapid environmental dilution of exogenous feedstock molecules, in the handful of local environments afforded by a single cometary impact.

\subsection*{Prospects for cometary delivery beyond Earth}

While our results demonstrate that cometary impacts were exceedingly unlikely to have played an important role in the origins of life on Earth -- the only known instance of life -- we find that cometary impacts will more readily deliver HCN, in high concentration, to Mars. Farther afield, many exoplanetary systems harbour large outer reservoirs of icy planetesimals in debris disks analogous to the Kuiper belt -- with many inferred to have masses 6 orders of magnitude larger\cite{KrivovWyatt2021}. Our results therefore clearly cannot preclude the possibility that, in favourable planetary contexts, cometary delivery may provide a robust source of concentrated prebiotic feedstock molecules.
Here, we discuss the implications of our results for the prospects of cometary delivery beyond Earth.

Comparison of the end-member cases of Mars and Venus, enabled by their shared dynamical origins, highlights several key planetary and planetary system requirements that control the efficiency, and thus importance, of cometary delivery in a general exoplanetary context. The significant differences between Mars and Venus are driven by atmospheric surface density, and the impact velocity distribution of short-period comets. The destruction of almost all comets smaller than $2$~km during descent through Venus' dense atmosphere, compounded by an average cometary impact velocity greater than $36~{\rm km\,s}^{-1}$\cite{Nesvorny2023}, restricts successful cometary delivery to an extremely small region of impact parameter space [\cite{McDonald2025}, Supplementary Text]. Mars, in contrast, benefits from its tenuous atmosphere that is expected to allow almost all ($D<100$~m) comets to reach the surface intact\cite{Anslow2025_atmos} with an average impact velocity of only $19.7~{\rm km\,s}^{-1}$\cite{Nesvorny2023}, which together supports the successful delivery of HCN in a much larger fraction of cometary impacts.

Given that the detailed atmospheric characterisation of terrestrial exoplanets remains elusive [e.g.,\cite{Greene2023:JWST_Trappist}], the key observationally accessible planetary property controlling the efficiency of cometary delivery is thus planet size. Current observational biases skew the present-day exoplanet population preferentially towards larger planets, such that well-characterised escape velocities are consistently, and substantially higher than of Earth (Supplementary Text). For this reason alone, we would expect cometary delivery to be significantly more challenging for the majority of terrestrial exoplanets discovered to-date.

Constraints on the efficiency of cometary delivery at the planetary system level are less straightforward, given the large number of (relevant) free parameters that would need to be considered on a system-by-system basis. Nevertheless, a minimum prerequisite is that the inner planetary system must experience a period of intense cometary bombardment in order to overcome the very low collision probability of highly-eccentric comets onto terrestrial planets with small semi-major axis [e.g.,\cite{Brasser2020,Nesvorny2023}]. Such a period of intense cometary bombardment is not, however, guaranteed in a general exoplanetary context, with the dynamical influence of exterior planets able to prevent, in the extreme case, the inwards transport of comets through the planetary system\cite{Wyatt2017:ScatteringDesign}. Crucially, this is most pronounced for low-mass stars around which even Earth-mass planets have the capacity to form effective dynamical barriers\cite{Wyatt2017:ScatteringDesign}.

We finally note that planetary system architecture and stellar mass both also play a significant role in determining the impact velocity distribution of comets onto terrestrial exoplanets. This is, again, most challenging around low-mass stars for which the minimum impact velocity of short-period comets is significantly higher than around Solar-type stars\cite{Anslow2023:DeliveryExoplanets}. This represents a significant, additional problem for cometary delivery given that the majority of Earth-like planets amenable to detailed atmospheric characterisation in the coming decades orbit low-mass M-dwarf stars -- the so-called ``M-dwarf opportunity''\cite{CharbonneauDeming2007}. 

It therefore appears overwhelmingly likely that cometary delivery will be substantially more challenging in a general exoplanetary context than onto Earth, driven by much higher characteristic cometary impact velocities and the efficient destruction of fragile prebiotic feedstock molecules. Taken together with our central result -- that Earth’s cometary bombardment was unable to provide the concentrated reservoir of prebiotic feedstock molecules required for the origins of life -- we expect the same to be true in a general exoplanetary context, challenging the long-held hypothesis that cometary impacts are likely to play an important role in the origins of life.

\section*{Methods}

\subsubsection*{The size-frequency distribution of Earth-crossing comets}

The dynamical simulations of Nesvorn\'y et al.\cite{Nesvorny2017:SPCs} require some physical mechanism to limit the number of perihelion passages of small comets to match the observed modern-day population of ecliptic comets. Given that the physical lifetimes of modern JFCs are an order of magnitude shorted than their dynamical lifetimes, and that observations demonstrate the frequent disruption of comets\cite{Sekanina1984,Boehnhardt2002,Jewitt2016}, this is primarily attributed to cometary fragmentation\cite{Nesvorny2017:SPCs}.

While the precise cause(s) of cometary fragmentation remain debated,
asymmetric outgassing has recently emerged as a leading
mechanism through which cometary nuclei can be spun-up to their
rotational disruption limits\cite{Jewitt2021,Safrit2021,Jewitt2022,LuuJewitt2024}. 
Indirect evidence supporting the dominant role of rotational disruption is provided, on a population level, by modern JFCs, which exhibit a `spin-barrier' in their radius-rotation period distribution, indicative of a population strongly sculpted by the effects of rotational instability\cite{Kokotanekova2017,Safrit2021,Jewitt2021}. 
This is further supported by the observed break-up of Comet 332P/Ikeya–Murakami\cite{Jewitt2016} and Comet 157P/Tritton\cite{LuuJewitt2024}, both consistent with rotational spin-up to the point of rotational instability. Finally, we note that the timescale of rotational spin-up is highly size-dependent [\cite{Jewitt2021}; see also Supplementary Text], satisfying the requirement for smaller comets to have shorter
physical lifetimes\cite{DiSisto2009,Nesvorny2017:SPCs}.

We follow a similar procedure to Rigley \& Wyatt\cite{RigleyWyatt2022} in numerically determining the size-frequency distribution (SFD) of Earth-crossing comets subject to rotational instability. 
We consider comets in the size range $0.1 - 1000$~km, divided into 80 logarithmically spaced bins, with initial SFD as they’re scattered into the inner Solar System represented by a broken power-law distribution, defined such that the number of comets of radius $R$ is given by
\begin{equation}
    n(R)dR = KR^{-\alpha}dR.
\end{equation}
The power law indices $\alpha$ are derived from available constraints from JFCs and the Kuiper belt (summarised in Table~\ref{tab:input_JFC_slopes}), while the constants of proportionality $K$ in each size-range are constrained via the continuity of the size distribution.

\begin{table}
    \centering
    \begin{tabular}{lcc}
        \hline
        Size range [km] & $\alpha$ \\
        \hline
        $0.1 \leq R \leq 1$ & $2.25$ \\
        $1 \leq R \leq 50$ & $3.0$ \\
        $50 \leq R \leq 150$ & $6.0$ \\
        $150 \leq R \leq 1000$ & $3.5$\\
        \hline
    \end{tabular}
    \caption{\textbf{Slopes of the differential size-frequency distribution (SFD) of Jupiter-family nuclei.}
    This observationally motivated distribution is used in our numerical model as the SFD of comets input into the inner Solar System. The slope of the cumulative size distribution, often quoted in the literature is $\gamma = \alpha - 1$.}
    \label{tab:input_JFC_slopes}
\end{table}

Comets are added from the source region every $12000$~yr -- the physical lifetime of modern JFCs\cite{LevisonDuncan1997} -- with the number of comets added to each size bin sampled randomly from a Poisson distribution with mean value given by the input SFD (Table~\ref{tab:input_JFC_slopes}). The probability a comet fragments per time-step ($\delta t$) is given by
\begin{equation}
    f(R) = \dfrac{\delta t}{\tau_{\rm ri}(R)},
\end{equation}
where
\begin{equation}
    \tau_{\rm ri}(R) = 100~\left(\dfrac{R}{1~{\rm km}}\right)^2~{\rm yr}
\end{equation}
is the (size-dependent) timescale of rotational instability [Jewitt\cite{Jewitt2021}; see Supplementary Text]. If fragmentation occurs, determined by sampling a random number in the range $[0, 1]$, it is assumed that two equal-mass child fragments are produced, distributed into the corresponding radius bin. 

This population of comets, subject to losses via rotational instability, is integrated for $10$~Myr (with a timestep $\delta t=1$~yr) until the SFD reaches a steady-state; which we ensure is insensitive to the choice of timestep. 
The strong dependence of physical lifetime on comet radius is evident, with the slope of the SFD significantly shallower than the source population at small sizes. For $0.1-1$~km comets, the slope of the cumulative SFD decreases from -1.25 to -0.78, and for $1-50$~km comets, the slope decreases from -2.0 to -1.34. The SFD at larger sizes is relatively unaltered, with the slope decreasing from -5.0 to -4.78 for comets in the size range $50-150$~km. 

We note that the resulting steady-state SFD is reassuringly insensitive to the assumed mechanism of comet fragmentation. Adopting instead the Di Sisto et al.\cite{DiSisto2009} model of comet fragmentation -- which is agnostic of physical break-up mechanism -- we recover a SFD with slopes -0.78, -1.34 and -4.74 respectively. This is in surprising agreement with the Jewitt\cite{Jewitt2021} model, and crucially, also implies a dramatic reduction in the number of small comets on Earth-crossing orbits.

\subsubsection*{Earth's cometary impact chronology}

Given the inability of the lunar crater record to constrain both the bombardment history of the Earth-Moon system pre-3.92~Gya\cite{Zellner2017} and impactor composition\cite{Bailey1991}, we instead leverage Earth's cometary bombardment chronology as derived from numerical N-body modelling of short-period comet formation and evolution\cite{Nesvorny2017:SPCs,Nesvorny2023}. 

We anchor Earth's cometary bombardment to the flux of $D>10$~km comets reported in Nesvorn\'y et al.\cite{Nesvorny2023}, given this subset of comets is least vulnerable to rotational instability. The cumulative flux of $D>10$~km comets is distributed in time according to
\begin{equation}
    F(D>10\,{\rm km}, t) = \dfrac{1}{6} \left[F_1 \exp{\left\{-\left(\dfrac{t}{\tau_1}\right)^{\alpha_1}\right\}}
    + F_2 \exp{\left\{-\left(\dfrac{t}{\tau_2}\right)^{\alpha_2}\right\}} + F_3 \left(4570 - t\right)\right],
\end{equation}
where $F_1=F_2=5.6\times10^3$, $\tau_1=7$~Myr, $\alpha_1=1$, $\tau_2=13$~Myr, $\alpha_2=0.44$, and $F_3=4\times10^{-3}$.

The impact flux of smaller comets is constrained by the steady-state SFD of Earth-crossing comets, and the continuity of the SFD at $D=10$~km,
\begin{equation}
    F(D, t) = F(D>10\,{\rm km}, t) \left\{\dfrac{K(D)}{K(D>10~{\rm km})}\right\},
\end{equation}
where
\begin{align}
    \nonumber K(D) = ~&D_{12}^{\alpha_1-\alpha_2}\int_{0.2}^{D_{12}}{D^{-\alpha_1}dD} + \int_{D_{12}}^{D_{23}}{D^{-\alpha_2}dD} +\\ &D_{23}^{\alpha_3 - \alpha_2} \int_{D_{23}}^{D_{34}}{D^{-\alpha_3}dD} + D_{34}^{\alpha_4 - \alpha_2}\int_{D_{34}}^{2000}{D^{-\alpha_4}dD},
\end{align}
and
\begin{equation}
    K(D>10~{\rm km}) = \int_{10}^{D_{23}}{D^{-\alpha_2}dD} + D_{23}^{\alpha_3 - \alpha_2} \int_{D_{23}}^{D_{34}}{D^{-\alpha_3}dD} + D_{34}^{\alpha_4 - \alpha_2}\int_{D_{34}}^{2000}{D^{-\alpha_4}dD}.
\end{equation}
Here, we have defined the transition diameters $D_{12}=2~$km, $D_{23}=100~$km, $D_{34}=300~$km, which relate to the power-law indices $\alpha_1 = 1.78$, $\alpha_2 = 2.34$, $\alpha_3 = 5.78$, $\alpha_4 = 3.5$ respectively.

\subsubsection*{The impact velocity distribution of short-period comets}

To quantify the cometary delivery of prebiotic feedstock molecules, we must specify the velocity distribution of incident comets, given this largely controls the survival of fragile feedstock molecules during hypervelocity impact\cite{PierazzoChyba1999,McDonald2025}.

Nesvorn\'y et al.\cite{Nesvorny2023} record the mean impact velocities of comets, using an {\" O}pik-like algorithm, onto the terrestrial planets, finding $[19.7, 28.2, 36.8]~{\rm km\,s}^{-1}$ onto Mars, Earth and Venus respectively. Motivated by the impact velocity distribution of meteoroids onto the modern Earth\cite{Rumpf2017,Drolshagen2020}, we sample impact velocities from a log-normal distribution,
\begin{equation}
    dF(v_{\rm imp}; \mu, \sigma) = \dfrac{1}{\sigma v_{\rm imp}\sqrt{2\pi}}\exp{\left(-\dfrac{(\ln{v_{\rm imp}} - \mu)^2}{2\sigma^2}\right)}dv_{\rm imp},
\end{equation}
where $\mu$ is the mean, and $\sigma$ the standard deviation of $\ln v_{\rm imp}$, respectively (not the arithmetic mean and standard deviation of the impact velocity itself).

Nesvorn\'y et al.\cite{Nesvorny2023} do not record the standard deviation of comet impact velocities; motivated by the results of alternative dynamical simulations of Earth's cometary bombardment\cite{Brasser2020}, and the modern velocity distribution of both short-period comets\cite{HughesWilliams2000} and small meteoroids\cite{Drolshagen2020}, we adopt an arithmetic velocity standard deviation of $s_v=5~{\rm km\,s^{-1}}$. The corresponding parameters of the log-normal distribution are therefore calculated as
\begin{align}
    \sigma^2 &= \ln\left[1+\left(\frac{s_v}{\bar{v}}\right)^2\right],\\
    \mu &= \ln(\bar{v})-\frac{\sigma^2}{2},
\end{align}
where $\bar{v}=28.2~{\rm km\,s^{-1}}$ and $s_v=5~{\rm km\,s^{-1}}$, such that the resulting distribution therefore has an arithmetic mean and standard deviation of $28.2$ and $5~{\rm km\,s^{-1}}$, respectively. For comparison, long-period comets have mean impact velocities in excess of $50~{\rm km\,s}^{-1}$\cite{HughesWilliams2000} -- substantially higher than that of short-period comets -- motivating our focus on the latter throughout the manuscript.

\subsubsection*{The entry angle distribution of comets}

Comets are assumed to arrive at the top of the atmosphere with a range of entry angles (defined relative to the horizontal) distributed according to the canonical Shoemaker\cite{Shoemaker1962} distribution,
\begin{equation}
    dF(\theta) = \sin{2\theta}d\theta.
\end{equation}
This distribution accounts for the gravitational focussing of comets by Earth, and corresponds to a most probable entry angle of $45^\circ$. Half of all cometary impacts will have entry angle in the range $30-60^\circ$, while less than 7\% of comets will arrive with entry angle shallower than $15^\circ$ or steeper than $75^\circ$.

\subsubsection*{The effects of atmospheric entry on Earth's cometary impact flux}

Given the low density and weak mechanical strength of cometary nuclei\cite{Groussin2019}, even relatively large comets will be significantly decelerated during atmospheric passage, while small comets will catastrophically fragment at high altitudes\cite{Chyba1993_tunguska,HillsGoda1993,Anslow2025_atmos}. A direct consequence of atmospheric entry is that the distribution of comets reaching Earth's surface will differ significantly from the population incident at the top of the atmosphere.

We use the \textsc{atmosentry} code\cite{Anslow2025_atmos} to model the interaction of comets with the atmosphere, and record the size, velocity, and angle of comets directly before hypervelocity impact. This numerical model, described in detail in Anslow et al.\cite{Anslow2025_atmos}, includes semi-analytical parametrisations for the ablation, deformation and fragmentation of comets during atmospheric entry. Key model parameters, consistent with previous literature [e.g.,\cite{Chyba1993_tunguska,HillsGoda1993,Anslow2025_atmos}] are described below.

Although geological constraints on the Hadean atmosphere are scarce and the details of its evolution remain debated, modelling suggests that it was initially very dense ($\sim100~{\rm bar}$) following the degassing of
Earth’s magma ocean\cite{Elkins-Tanton2008}. A relatively quick transition is then expected to $\lesssim1.1~{\rm bar}$ during the Archean\cite{CatlingZahnle2020}. In the interests of simplicity, we thus assume an isothermal atmospheric with density profile
\begin{equation}
    \rho_{\rm atm}(z) = \rho_{\rm surf} \exp{\left(-\dfrac{z}{H}\right)},
\end{equation}
with scale height $H=7.2$~km and surface density $\rho_{\rm surf}=1.225~{\rm kg\,m}^{-3}$, roughly consistent with Earth's modern atmosphere.
We assume a cometary bulk density of $0.6~{\rm g\,cm}^{-3}$\cite{Kokotanekova2017} and tensile strength $10^4~$Pa\cite{Groussin2019}. These are both upper limits, ensuring our results will present a best-case scenario for the cometary delivery of prebiotic feedstock molecules. 
In particular, sensitivity tests reported in\cite{Anslow2025_atmos} demonstrate that lower bulk densities would reduce the minimum comet size able to survive atmospheric entry.

\subsubsection*{Exposed land on the early Earth}

The dilution of prebiotic feedstock molecules during cometary impacts into global oceans presents a major concern, with only impacts onto subaerial land able to deliver a prebiotically plausible concentration of HCN. The extent of exposed land on the early Earth remains, however, highly uncertain\cite{Korenaga2021}.
Moreover, while there exist many models of continental growth, these are crucially not direct proxies for exposed landmass given the non-monotonic history of ocean volume [\cite{Korenaga2021}; see Supplementary Text for further discussion].

For simplicity, given the significant uncertainty in estimates of exposed land on the early Earth, we adopt the McCulloch \& Bennett\cite{McCullochBennett1993} model of continental growth as a proxy for exposed landmass (following\cite{Pearce2017}). As noted in Bada \& Korenaga\cite{BadaKorenaga2018}, this is likely a large overestimate,
which guarantees that our results represent an upper limit on the number of successful cometary impacts onto the early Earth. In this model, crustal surface area increases linearly from $0$ to $12.8$\% of the present-day crustal surface area between $4.5-3.7$~Gya, such that the fraction of exposed land is given by
\begin{equation}
    f_{\rm cc} = 0.0656~t,
\end{equation}
where $t$ is in Gyr.

While there remains significant debate regarding the growth of continental crust, there exists consensus with respect to the limited extent of continental crust during the early Hadean. Indeed, it is plausible that exposed land was limited to hotspot oceanic islands\cite{Korenaga2021,BadaKorenaga2018}. Given the rapid decline in Earth's cometary bombardment\cite{Nesvorny2023}, the divergence of these models at later times has very little influence on our results; adopting instead a model that favours much quicker crustal formation\cite{GuoKorenaga2020}, the number of successful impacts onto Earth increases by only a factor of $3$ (Supplementary Text).

\subsubsection*{Monte Carlo simulation of Earth's cometary bombardment}

To quantitatively investigate the successful cometary delivery of prebiotic feedstock molecules we employ Monte Carlo simulation, discretising Earth's cometary impact flux into 1~Myr timesteps. 
At each timestep, $t_i$, the expected number of cometary impacts onto exposed land, $\bar{N}_i$, is given by
\begin{equation}
    \bar{N}_i = [F(D>100~{\rm m}, t_i) - F(D>100~{\rm m}, t_{i+1})] f_{\rm cc}(t_i),
\end{equation}
where, recall, $F(D>100~{\rm m}, t_i)$ is the cumulative number of $D>100$~m cometary impacts at $t_i$, and $f_{\rm cc}(t_i)$ is the available exposed landmass. This is, generally, not an integer; the number of cometary impacts, $N_i$, at each timestep is thus sampled from a Poisson distribution with mean value $\bar{N}_i$. We note that $D<100$~m comets will catastrophically airburst at high altitude\cite{Anslow2025_atmos}, which allows us to limit the number of number of comets sampled, and thereby greatly decrease the computational cost of these simulations.

At each timestep we generate comets' radii, velocities, and entry angles at the top-of-atmosphere from the analytical distributions described previously. The effects of atmospheric entry are then calculated for each comet, determining the final distribution of radii, velocities, and angles at the surface. Not all comets will survive atmospheric entry, depending primarily on their initial size and entry angle; the number of comets reaching the surface will therefore be less than $N_i$. 

For the subset of comets surviving atmospheric entry, we finally determine the fraction of HCN surviving hypervelocity impact. Comets are assumed to have an initial HCN concentration of $0.1$~\% level relative to water, consistent with Solar System comets\cite{MummaCharnley2011}, and we use the parametrisation presented in McDonald et al.\cite{McDonald2025} (Eqs.~14-15), assuming equilibrium chemistry. As the HCN survival parametrisation presented in McDonald et al.\cite{McDonald2025} does not extend to comets smaller than $100~{\rm m}$, or to impact angles shallower than $15^\circ$ (relative to the horizontal), we discard any comets, or surviving fragments, meeting these criteria.

This corresponds to an extremely small region of parameter space; less than $7$\% of comets arrive with entry angle less than $15^\circ$, with only much larger (roughly km-scale) comets able to survive atmospheric entry due to the increased atmospheric path\cite{Anslow2025_atmos}. Moreover, atmospheric entry acts to increase impact angles towards the normal (see Fig.~\ref{fig:TOA_surf_dists_terrestrial}); this restriction therefore affects only km-scale comets with entry angles less than $15^\circ$. Finally, as discussed in\cite{Anslow2025_atmos}, only a small subset of comets will generate $D<100$~m fragments that reach Earth's surface; these fragments are expected to be highly thermally processed, and thus constitute an unlikely source of concentrated HCN. We therefore do not expect this exclusion to materially affect our key conclusions, given that successful delivery of HCN would in any case require a low-velocity impact, further reducing this subset of comets. Nonetheless, we note that it remains possible in principle for a highly oblique, low-velocity comet to deliver a prebiotically relevant concentration of HCN. We address this possibility via a heuristic toy model in the Discussion (see Fig.~\ref{eq:origins_of_life_timescale}), and note that some fraction of our simulations do, regardless, record a small number of successful cometary impacts delivering $>10~{\rm mM}$ HCN.

To account for the inherent stochasticity when sampling from the respective distributions, we repeat this procedure $10^5$ times, recording the number of successful cometary impacts as a function of time.

\section*{Funding declaration}

R.J.A. acknowledges the Science and Technology Facilities Council (STFC) for a PhD studentship. 
C.H.M. gratefully acknowledges the Leverhulme Centre for Life in the Universe at the University of Cambridge for support through Joint Collaborations Research Project Grant GAG/382.
A.B. acknowledges the support of a Royal Society University Research Fellowship, URF/R1/211421.

\bibliography{sample}

@Misc{methods,
  note = {Materials and methods are available as supplementary material},
}

@article{Anslow2023:DeliveryExoplanets,
    author      =   {{Anslow}, R. J. and {Bonsor}, A. and {Rimmer}, P. B.},
    title       =   "{Can comets deliver prebiotic molecules to rocky exoplanets?}",
    journal     =   {Proc. R. Soc. Lond., A},
    volume      =   {479},
    number      =   {2279},
    pages       =   {20230434},
    year        =   {2023},
    doi         =   {10.1098/rspa.2023.0434}
}

@ARTICLE{Anslow2025_overlap,
       author = {{Anslow}, R.~J. and {Bonsor}, A. and {Rimmer}, P.~B. and {Rae}, A.~S.~P. and {McDonald}, C.~H. and {Walton}, C.~R.},
        title = "{The plausibility of origins scenarios requiring two impactors}",
      journal = {Proceedings of the Royal Society of London Series A},
         year = 2025,
        month = jan,
       volume = {481},
          eid = {20240327},
        pages = {20240327},
          doi = {10.1098/rspa.2024.0327},
       adsurl = {https://ui.adsabs.harvard.edu/abs/2025RSPSA.48140327A}
}

@ARTICLE{Anslow2025_atmos,
       author = {{Anslow}, Richard J. and {Bonsor}, Amy and {Todd}, Zoe R. and {Wordsworth}, Robin and {Rae}, Auriol S.~P. and {McDonald}, Catriona H. and {Rimmer}, Paul B.},
        title = "{The atmospheric entry of cometary impactors}",
      journal = {Monthly Notices of the Royal Astronomical Society},
         year = 2025,
        month = may,
       volume = {539},
       number = {1},
        pages = {376-392},
          doi = {10.1093/mnras/staf507},
       adsurl = {https://ui.adsabs.harvard.edu/abs/2025MNRAS.539..376A}
}

@ARTICLE{AulbachStagno2016,
       author = {{Aulbach}, Sonja and {Stagno}, Vincenzo},
        title = "{Evidence for a reducing Archean ambient mantle and its effects on the carbon cycle}",
      journal = {Geology},
         year = 2016,
        month = sep,
       volume = {44},
       number = {9},
        pages = {751-754},
          doi = {10.1130/G38070.1},
       adsurl = {https://ui.adsabs.harvard.edu/abs/2016Geo....44..751A}
}

@ARTICLE{BadaKorenaga2018,
       author = {{Bada}, Jeffrey L. and {Korenaga}, Jun},
        title = "{Exposed Areas Above Sea Level on Earth >3.5 Gyr Ago: Implications for Prebiotic and Primitive Biotic Chemistry}",
      journal = {Life},
         year = 2018,
        month = nov,
       volume = {8},
       number = {4},
          eid = {55},
        pages = {55},
          doi = {10.3390/life8040055},
       adsurl = {https://ui.adsabs.harvard.edu/abs/2018Life....8...55B}
}

@ARTICLE{Bailey1991,
       author = {{Bailey}, M.~E.},
        title = "{Comet craters versus asteroid craters}",
      journal = {Advances in Space Research},
         year = 1991,
        month = jan,
       volume = {11},
       number = {6},
        pages = {43-60},
          doi = {10.1016/0273-1177(91)90231-8}
}

@ARTICLE{Becker2019,
       author = {{Becker}, Sidney and {Feldmann}, Jonas and {Wiedemann}, Stefan and {Okamura}, Hidenori and {Schneider}, Christina and {Iwan}, Katharina and {Crisp}, Antony and {Rossa}, Martin and {Amatov}, Tynchtyk and {Carell}, Thomas},
        title = "{Unified prebiotically plausible synthesis of pyrimidine and purine RNA ribonucleotides}",
      journal = {Science},
         year = 2019,
        month = oct,
       volume = {366},
       number = {6461},
        pages = {76-82},
          doi = {10.1126/science.aax2747},
       adsurl = {https://ui.adsabs.harvard.edu/abs/2019Sci...366...76B}
}

@article{Benner2020,
    author = {Benner, Steven A. and Bell, Elizabeth A. and Biondi, Elisa and Brasser, Ramon and Carell, Thomas and Kim, Hyo-Joong and Mojzsis, Stephen J. and Omran, Arthur and Pasek, Matthew A. and Trail, Dustin},
    title = {When Did Life Likely Emerge on Earth in an RNA-First Process?},
    journal = {ChemSystemsChem},
    volume = {2},
    number = {2},
    pages = {e1900035},
    doi = {https://doi.org/10.1002/syst.201900035},
    url = {https://chemistry-europe.onlinelibrary.wiley.com/doi/abs/10.1002/syst.201900035},
    eprint = {https://chemistry-europe.onlinelibrary.wiley.com/doi/pdf/10.1002/syst.201900035},
    year = {2020}
}

@ARTICLE{Boehnhardt2002,
       author = {{Boehnhardt}, Hermann},
        title = "{Comet Splitting - Observations and Model Scenarios}",
      journal = {Earth Moon and Planets},
         year = 2002,
        month = oct,
       volume = {89},
       number = {1},
        pages = {91-115},
          doi = {10.1023/A:1021538201389},
       adsurl = {https://ui.adsabs.harvard.edu/abs/2002EM&P...89...91B}
    }

@ARTICLE{Brasser2020,
       author = {{Brasser}, R. and {Werner}, S.~C. and {Mojzsis}, S.~J.},
        title = "{Impact bombardment chronology of the terrestrial planets from 4.5 Ga to 3.5 Ga}",
      journal = {Icarus},
         year = 2020,
        month = mar,
       volume = {338},
          eid = {113514},
        pages = {113514},
          doi = {10.1016/j.icarus.2019.113514},
       adsurl = {https://ui.adsabs.harvard.edu/abs/2020Icar..33813514B}
}

@article{Cassata2025,
    title = {A refined isotopic composition of cometary xenon and implications for the accretion of comets and carbonaceous chondrites on Earth},
    journal = {Earth and Planetary Science Letters},
    volume = {660},
    pages = {119307},
    year = {2025},
    issn = {0012-821X},
    doi = {https://doi.org/10.1016/j.epsl.2025.119307},
    url = {https://www.sciencedirect.com/science/article/pii/S0012821X25001062},
    author = {William S. Cassata},
}

@ARTICLE{Carr1987,
       author = {{Carr}, M.~H.},
        title = "{Water on Mars}",
      journal = {Nature},
         year = 1987,
        month = mar,
       volume = {326},
       number = {6108},
        pages = {30-35},
          doi = {10.1038/326030a0},
       adsurl = {https://ui.adsabs.harvard.edu/abs/1987Natur.326...30C}
}

@BOOK{CatlingKasting2017,
       author = {{Catling}, David C. and {Kasting}, James F.},
        title = "{Atmospheric Evolution on Inhabited and Lifeless Worlds}",
         year = 2017,
       adsurl = {https://ui.adsabs.harvard.edu/abs/2017aeil.book.....C}
}

@ARTICLE{CatlingZahnle2020,
       author = {{Catling}, David C. and {Zahnle}, Kevin J.},
        title = "{The Archean atmosphere}",
      journal = {Science Advances},
         year = 2020,
        month = feb,
       volume = {6},
       number = {9},
          eid = {eaax1420},
        pages = {eaax1420},
          doi = {10.1126/sciadv.aax1420},
       adsurl = {https://ui.adsabs.harvard.edu/abs/2020SciA....6.1420C}
}

@article{ChamberlinChamberlin1908,
    author = {T. C. Chamberlin  and R. T. Chamberlin },
    title = {Early Terrestrial Conditions That May Have Favored Organic Synthesis},
    journal = {Science},
    volume = {28},
    number = {730},
    pages = {897-911},
    year = {1908},
    doi = {10.1126/science.28.730.897},
}

@ARTICLE{CharbonneauDeming2007,
       author = {{Charbonneau}, David and {Deming}, Drake},
        title = "{The Dynamics-Based Approach to Studying Terrestrial Exoplanets}",
      journal = {arXiv e-prints},
         year = 2007,
        month = jun,
          eid = {arXiv:0706.1047},
        pages = {arXiv:0706.1047},
          doi = {10.48550/arXiv.0706.1047},
archivePrefix = {arXiv},
       eprint = {0706.1047},
 primaryClass = {astro-ph},
       adsurl = {https://ui.adsabs.harvard.edu/abs/2007arXiv0706.1047C}
}

@ARTICLE{ChenKipping2018,
       author = {{Chen}, Jingjing and {Kipping}, David},
        title = "{On the Rate of Abiogenesis from a Bayesian Informatics Perspective}",
      journal = {Astrobiology},
         year = 2018,
        month = dec,
       volume = {18},
       number = {12},
        pages = {1574-1584},
          doi = {10.1089/ast.2018.1836},
archivePrefix = {arXiv},
       eprint = {1806.08033},
 primaryClass = {astro-ph.EP},
       adsurl = {https://ui.adsabs.harvard.edu/abs/2018AsBio..18.1574C}
}

@article{Chyba1990,
     ISSN = {00368075, 10959203},
     author = {Christopher F. Chyba and Paul J. Thomas and Leigh Brookshaw and Carl Sagan},
     journal = {Science},
     number = {4967},
     pages = {366--373},
     publisher = {American Association for the Advancement of Science},
     title = {Cometary Delivery of Organic Molecules to the Early Earth},
     urldate = {2023-03-11},
     volume = {249},
     year = {1990}
}

@ARTICLE{ChybaSagan1992,
       author = {{Chyba}, Christopher and {Sagan}, Carl},
        title = "{Endogenous production, exogenous delivery and impact-shock synthesis of organic molecules: an inventory for the origins of life}",
      journal = {Nature},
         year = 1992,
        month = jan,
       volume = {355},
       number = {6356},
        pages = {125-132},
          doi = {10.1038/355125a0},
       adsurl = {https://ui.adsabs.harvard.edu/abs/1992Natur.355..125C}
}

@ARTICLE{Chyba1993_tunguska,
       author = {{Chyba}, C.~F. and {Thomas}, P.~J. and {Zahnle}, K.~J.},
        title = "{The 1908 Tunguska explosion: atmospheric disruption of a stony asteroid}",
      journal = {Nature},
         year = 1993,
        month = jan,
       volume = {361},
       number = {6407},
        pages = {40-44},
          doi = {10.1038/361040a0},
       adsurl = {https://ui.adsabs.harvard.edu/abs/1993Natur.361...40C}
}

@article{Clark1988,
	Author = {Clark, Benton C. },
	Da = {1988/09/01},
	Doi = {10.1007/BF01804671},
	Id = {Clark1988},
	Isbn = {1573-0875},
	Journal = {Origins of life and evolution of the biosphere},
	Number = {3},
	Pages = {209--238},
	Title = {Primeval procreative comet pond},
	Ty = {JOUR},
	Volume = {18},
	Year = {1988}
}

@ARTICLE{Clemett1993,
       author = {{Clemett}, Simon J. and {Maechling}, Claude R. and {Zare}, Richard N. and {Swan}, Patrick D. and {Walker}, Robert M.},
        title = "{Identification of Complex Aromatic Molecules in Individual Interplanetary Dust Particles}",
      journal = {Science},
         year = 1993,
        month = oct,
       volume = {262},
       number = {5134},
        pages = {721-725},
          doi = {10.1126/science.262.5134.721}
}

@ARTICLE{Constantinou2025:DryVenus,
       author = {{Constantinou}, T. and {Shorttle}, O. and {Rimmer}, P. B.},
        title = "{A dry Venusian interior constrained by atmospheric chemistry}",
      journal = {Nature Astronomy},
         year = 2025,
        month = feb,
       volume = {9},
        pages = {189-198},
          doi = {10.1038/s41550-024-02414-5}
}

@ARTICLE{DiSisto2009,
       author = {{Di Sisto}, Romina P. and {Fern{\'a}ndez}, Julio A. and {Brunini}, Adri{\'a}n},
        title = "{On the population, physical decay and orbital distribution of Jupiter family comets: Numerical simulations}",
      journal = {Icarus},
         year = 2009,
        month = sep,
       volume = {203},
       number = {1},
        pages = {140-154},
          doi = {10.1016/j.icarus.2009.05.002}
}

@ARTICLE{Drolshagen2020,
       author = {{Drolshagen}, Esther and {Ott}, Theresa and {Koschny}, Detlef and {Drolshagen}, Gerhard and {Schmidt}, Anna Kristiane and {Poppe}, Bj{\"o}rn},
        title = "{Velocity distribution of larger meteoroids and small asteroids impacting Earth}",
      journal = {Planetary and Space Science},
         year = 2020,
        month = may,
       volume = {184},
          eid = {104869},
        pages = {104869},
          doi = {10.1016/j.pss.2020.104869}
}

@ARTICLE{Elkins-Tanton2008,
       author = {{Elkins-Tanton}, L.~T.},
        title = "{Linked magma ocean solidification and atmospheric growth for Earth and Mars}",
      journal = {Earth and Planetary Science Letters},
         year = 2008,
        month = jul,
       volume = {271},
       number = {1-4},
        pages = {181-191},
          doi = {10.1016/j.epsl.2008.03.062}
}

@ARTICLE{Foden2020,
       author = {{Foden}, Callum S. and {Islam}, Saidul and {Fern{\'a}ndez-Garc{\'\i}a}, Christian and {Maugeri}, Leonardo and {Sheppard}, Tom D. and {Powner}, Matthew W.},
        title = "{Prebiotic synthesis of cysteine peptides that catalyze peptide ligation in neutral water}",
      journal = {Science},
         year = 2020,
        month = nov,
       volume = {370},
       number = {6518},
        pages = {865-869},
          doi = {10.1126/science.abd5680}
}

@ARTICLE{Greene2023:JWST_Trappist,
       author = {{Greene}, Thomas P. and {Bell}, Taylor J. and {Ducrot}, Elsa and {Dyrek}, Achr{\`e}ne and {Lagage}, Pierre-Olivier and {Fortney}, Jonathan J.},
        title = "{Thermal emission from the Earth-sized exoplanet TRAPPIST-1 b using JWST}",
      journal = {Nature},
         year = 2023,
        month = jun,
       volume = {618},
       number = {7963},
        pages = {39-42},
          doi = {10.1038/s41586-023-05951-7}
    }

@ARTICLE{Groussin2019,
       author = {{Groussin}, O. and {Attree}, N. and {Brouet}, Y. and {Ciarletti}, V. and {Davidsson}, B. and {Filacchione}, G. and {Fischer}, H. -H. and {Gundlach}, B. and {Knapmeyer}, M. and {Knollenberg}, J. and {Kokotanekova}, R. and {K{\"u}hrt}, E. and {Leyrat}, C. and {Marshall}, D. and {Pelivan}, I. and {Skorov}, Y. and {Snodgrass}, C. and {Spohn}, T. and {Tosi}, F.},
        title = "{The Thermal, Mechanical, Structural, and Dielectric Properties of Cometary Nuclei After Rosetta}",
      journal = {Space Science Reviews},
         year = 2019,
        month = jun,
       volume = {215},
       number = {4},
          eid = {29},
        pages = {29},
          doi = {10.1007/s11214-019-0594-x},
       adsurl = {https://ui.adsabs.harvard.edu/abs/2019SSRv..215...29G}
}

@ARTICLE{GuoKorenaga2020,
       author = {{Guo}, Meng and {Korenaga}, Jun},
        title = "{Argon constraints on the early growth of felsic continental crust}",
      journal = {Science Advances},
         year = 2020,
        month = may,
       volume = {6},
       number = {21},
        pages = {eaaz6234},
          doi = {10.1126/sciadv.aaz6234},
       adsurl = {https://ui.adsabs.harvard.edu/abs/2020SciA....6.6234G}
}

@ARTICLE{Hanni2021,
       author = {{H{\"a}nni}, N. and {Altwegg}, K. and {Balsiger}, H. and {Combi}, M. and {Fuselier}, S.~A. and {De Keyser}, J. and {Pestoni}, B. and {Rubin}, M. and {Wampfler}, S.~F.},
        title = "{Cyanogen, cyanoacetylene, and acetonitrile in comet 67P and their relation to the cyano radical}",
      journal = {Astronomy \& Astrophysics},
         year = 2021,
        month = mar,
       volume = {647},
          eid = {A22},
        pages = {A22},
          doi = {10.1051/0004-6361/202039580},
archivePrefix = {arXiv},
       eprint = {2102.03433},
 primaryClass = {astro-ph.EP},
       adsurl = {https://ui.adsabs.harvard.edu/abs/2021A&A...647A..22H}
}

@ARTICLE{HillsGoda1993,
       author = {{Hills}, J.~G. and {Goda}, M.~P.},
        title = "{The Fragmentation of Small Asteroids in the Atmosphere}",
      journal = {The Astronomical Journal},
         year = 1993,
        month = mar,
       volume = {105},
        pages = {1114},
          doi = {10.1086/116499},
       adsurl = {https://ui.adsabs.harvard.edu/abs/1993AJ....105.1114H}
}

@ARTICLE{HughesWilliams2000,
       author = {{Hughes}, David W. and {Williams}, Iwan P.},
        title = "{The velocity distributions of periodic comets and stream meteoroids}",
      journal = {Monthly Notices of the Royal Astronomical Society},
         year = 2000,
        month = jul,
       volume = {315},
       number = {3},
        pages = {629-634},
          doi = {10.1046/j.1365-8711.2000.03435.x},
       adsurl = {https://ui.adsabs.harvard.edu/abs/2000MNRAS.315..629H}
}

@ARTICLE{Jewitt2016,
       author = {{Jewitt}, David and {Mutchler}, Max and {Weaver}, Harold and {Hui}, Man-To and {Agarwal}, Jessica and {Ishiguro}, Masateru and {Kleyna}, Jan and {Li}, Jing and {Meech}, Karen and {Micheli}, Marco and {Wainscoat}, Richard and {Weryk}, Robert},
        title = "{Fragmentation Kinematics in Comet 332P/Ikeya-Murakami}",
      journal = {The Astrophysical Journal Letters},
         year = 2016,
        month = sep,
       volume = {829},
       number = {1},
          eid = {L8},
        pages = {L8},
          doi = {10.3847/2041-8205/829/1/L8},
       adsurl = {https://ui.adsabs.harvard.edu/abs/2016ApJ...829L...8J}
}

@ARTICLE{Jewitt2021,
       author = {{Jewitt}, David},
        title = "{Systematics and Consequences of Comet Nucleus Outgassing Torques}",
      journal = {The Astronomical Journal},
         year = 2021,
        month = jun,
       volume = {161},
       number = {6},
          eid = {261},
        pages = {261},
          doi = {10.3847/1538-3881/abf09c},
       adsurl = {https://ui.adsabs.harvard.edu/abs/2021AJ....161..261J}
}

@ARTICLE{Jewitt2022,
       author = {{Jewitt}, David},
        title = "{Destruction of Long-period Comets}",
      journal = {The Astronomical Journal},
         year = 2022,
        month = oct,
       volume = {164},
       number = {4},
          eid = {158},
        pages = {158},
          doi = {10.3847/1538-3881/ac886d},
       adsurl = {https://ui.adsabs.harvard.edu/abs/2022AJ....164..158J}
}

@ARTICLE{Kokotanekova2017,
       author = {{Kokotanekova}, R. and {Snodgrass}, C. and {Lacerda}, P. and {Green}, S.~F. and {Lowry}, S.~C. and {Fern{\'a}ndez}, Y.~R. and {Tubiana}, C. and {Fitzsimmons}, A. and {Hsieh}, H.~H.},
        title = "{Rotation of cometary nuclei: new light curves and an update of the ensemble properties of Jupiter-family comets}",
      journal = {Monthly Notices of the Royal Astronomical Society},
         year = 2017,
        month = nov,
       volume = {471},
       number = {3},
        pages = {2974-3007},
          doi = {10.1093/mnras/stx1716},
       adsurl = {https://ui.adsabs.harvard.edu/abs/2017MNRAS.471.2974K}
}

@ARTICLE{Korenaga2017,
       author = {{Korenaga}, Jun and {Planavsky}, Noah J. and {Evans}, David A.~D.},
        title = "{Global water cycle and the coevolution of the Earth's interior and surface environment}",
      journal = {Philosophical Transactions of the Royal Society of London Series A},
         year = 2017,
        month = apr,
       volume = {375},
       number = {2094},
          eid = {20150393},
        pages = {20150393},
          doi = {10.1098/rsta.2015.0393},
       adsurl = {https://ui.adsabs.harvard.edu/abs/2017RSPTA.37550393K}
}

@ARTICLE{Korenaga2021,
       author = {{Korenaga}, Jun},
        title = "{Was There Land on the Early Earth?}",
      journal = {Life},
         year = 2021,
        month = oct,
       volume = {11},
       number = {11},
          eid = {1142},
        pages = {1142},
          doi = {10.3390/life11111142},
       adsurl = {https://ui.adsabs.harvard.edu/abs/2021Life...11.1142K}
}

@ARTICLE{KrivovWyatt2021,
       author = {{Krivov}, Alexander V. and {Wyatt}, Mark C.},
        title = "{Solution to the debris disc mass problem: planetesimals are born small?}",
      journal = {Monthly Notices of the Royal Astronomical Society},
         year = 2021,
        month = jan,
       volume = {500},
       number = {1},
        pages = {718-735},
          doi = {10.1093/mnras/staa2385},
archivePrefix = {arXiv},
       eprint = {2008.07406},
 primaryClass = {astro-ph.EP},
       adsurl = {https://ui.adsabs.harvard.edu/abs/2021MNRAS.500..718K}
}

@ARTICLE{LevisonDuncan1997,
       author = {{Levison}, Harold F. and {Duncan}, Martin J.},
        title = "{From the Kuiper Belt to Jupiter-Family Comets: The Spatial Distribution of Ecliptic Comets}",
      journal = {Icarus},
         year = 1997,
        month = may,
       volume = {127},
       number = {1},
        pages = {13-32},
          doi = {10.1006/icar.1996.5637}
}

@article{LuuJewitt2024,
    doi = {10.3847/1538-3881/ad8e62},
    url = {https://dx.doi.org/10.3847/1538-3881/ad8e62},
    year = {2024},
    month = {dec},
    publisher = {The American Astronomical Society},
    volume = {169},
    number = {1},
    pages = {14},
    author = {Jane Luu and David Jewitt},
    title = {Continuous Fragmentation of Comet 157P/Tritton},
    journal = {The Astronomical Journal},
}

@ARTICLE{Marty2017,
       author = {{Marty}, B. and {Altwegg}, K. and {Balsiger}, H. and {Bar-Nun}, A. and {Bekaert}, D.~V. and {Berthelier}, J. -J. and {Bieler}, A. and {Briois}, C. and {Calmonte}, U. and {Combi}, M. and {De Keyser}, J. and {Fiethe}, B. and {Fuselier}, S.~A. and {Gasc}, S. and {Gombosi}, T.~I. and {Hansen}, K.~C. and {H{\"a}ssig}, M. and {J{\"a}ckel}, A. and {Kopp}, E. and {Korth}, A. and {Le Roy}, L. and {Mall}, U. and {Mousis}, O. and {Owen}, T. and {R{\`e}me}, H. and {Rubin}, M. and {S{\'e}mon}, T. and {Tzou}, C. -Y. and {Waite}, J.~H. and {Wurz}, P.},
        title = "{Xenon isotopes in 67P/Churyumov-Gerasimenko show that comets contributed to Earth's atmosphere}",
      journal = {Science},
         year = 2017,
        month = jun,
       volume = {356},
       number = {6342},
        pages = {1069-1072},
          doi = {10.1126/science.aal3496},
       adsurl = {https://ui.adsabs.harvard.edu/abs/2017Sci...356.1069M}
}

@ARTICLE{McLennan2019,
       author = {{McLennan}, Scott M. and {Grotzinger}, John P. and {Hurowitz}, Joel A. and {Tosca}, Nicholas J.},
        title = "{The Sedimentary Cycle on Early Mars}",
      journal = {Annual Review of Earth and Planetary Sciences},
         year = 2019,
        month = may,
       volume = {47},
        pages = {91-118},
          doi = {10.1146/annurev-earth-053018-060332},
       adsurl = {https://ui.adsabs.harvard.edu/abs/2019AREPS..47...91M}
}

@ARTICLE{McCullochBennett1993,
       author = {{McCulloch}, Malcolm T. and {Bennett}, Victoria C.},
        title = "{Evolution of the early Earth: Constraints from $^{143}$Nd $^{142}$Nd isotopic systematics}",
      journal = {Lithos},
         year = 1993,
        month = sep,
       volume = {30},
       number = {3},
        pages = {237-255},
          doi = {10.1016/0024-4937(93)90038-E},
       adsurl = {https://ui.adsabs.harvard.edu/abs/1993Litho..30..237M}
}

@article{McDonald2025,
    title = {Constraining the survival of HCN during cometary impacts},
    journal = {Icarus},
    pages = {116704},
    year = {2025},
    issn = {0019-1035},
    doi = {https://doi.org/10.1016/j.icarus.2025.116704},
    author = {Catriona H. McDonald and Amy Bonsor and Auriol S.P. Rae and Paul B. Rimmer and Richard J. Anslow and Zoe R. Todd}
}

@BOOK{Melosh1989:ImpactCratering,
       author = {{Melosh}, H.~J.},
        title = "{Impact cratering : a geologic process}",
         year = 1989,
    publisher = {Oxford University Press}
}

@ARTICLE{Messager2016,
       author = {{Messager}, Mathis Lo{\"\i}c and {Lehner}, Bernhard and {Grill}, G{\"u}nther and {Nedeva}, Irena and {Schmitt}, Oliver},
        title = "{Estimating the volume and age of water stored in global lakes using a geo-statistical approach}",
      journal = {Nature Communications},
         year = 2016,
        month = dec,
       volume = {7},
          eid = {13603},
        pages = {13603},
          doi = {10.1038/ncomms13603},
       adsurl = {https://ui.adsabs.harvard.edu/abs/2016NatCo...713603M}
}

@ARTICLE{Miyakawa2002,
       author = {{Miyakawa}, Shin and {James Cleaves}, H. and {Miller}, Stanley L.},
        title = "{The Cold Origin of Life: A. Implications Based On The Hydrolytic Stabilities Of Hydrogen Cyanide And Formamide}",
      journal = {Origins of Life and Evolution of the Biosphere},
         year = 2002,
        month = jun,
       volume = {32},
       number = {3},
        pages = {195-208},
          doi = {10.1023/A:1016514305984},
       adsurl = {https://ui.adsabs.harvard.edu/abs/2002OLEB...32..195M}
}

@ARTICLE{Moody2024,
  title    = "The nature of the last universal common ancestor and its impact on the early Earth system",
  author   = "Moody, Edmund R R and {\'A}lvarez-Carretero, Sandra and Mahendrarajah, Tara A and Clark, James W and Betts, Holly C and Dombrowski, Nina and Sz{\'a}nth{\'o}, L{\'e}n{\'a}rd L and Boyle, Richard A and Daines, Stuart and Chen, Xi and Lane, Nick and Yang, Ziheng and Shields, Graham A and Sz{\"o}ll{\H o}si, Gergely J and Spang, Anja and Pisani, Davide and Williams, Tom A and Lenton, Timothy M and Donoghue, Philip C J",
  journal  = "Nature Ecology \& Evolution",
  volume   =  8,
  number   =  9,
  pages    = "1654--1666",
  month    =  sep,
  year     =  2024
}

@ARTICLE{MummaCharnley2011,
       author = {{Mumma}, Michael J. and {Charnley}, Steven B.},
        title = "{The Chemical Composition of Comets{\textemdash}Emerging Taxonomies and Natal Heritage}",
      journal = {Annual Review of Astronomy and Astrophysics},
         year = 2011,
        month = sep,
       volume = {49},
       number = {1},
        pages = {471-524},
          doi = {10.1146/annurev-astro-081309-130811}
}

@ARTICLE{Nesvorny2017:SPCs,
       author = {{Nesvorn{\'y}}, David and {Vokrouhlick{\'y}}, David and {Dones}, Luke and {Levison}, Harold F. and {Kaib}, Nathan and {Morbidelli}, Alessandro},
        title = "{Origin and Evolution of Short-period Comets}",
      journal = {The Astrophysical Journal},
         year = 2017,
        month = aug,
       volume = {845},
       number = {1},
          eid = {27},
        pages = {27},
          doi = {10.3847/1538-4357/aa7cf6}
}

@ARTICLE{Nesvorny2019,
       author = {{Nesvorn{\'y}}, David and {Vokrouhlick{\'y}}, David and {Stern}, Alan S. and {Davidsson}, Bj{\"o}rn and {Bannister}, Michele T. and {Volk}, Kathryn and {Chen}, Ying-Tung and {Gladman}, Brett J. and {Kavelaars}, J.~J. and {Petit}, Jean-Marc and {Gwyn}, Stephen D.~J. and {Alexandersen}, Mike},
        title = "{OSSOS. XIX. Testing Early Solar System Dynamical Models Using OSSOS Centaur Detections}",
      journal = {The Astronomical Journal},
         year = 2019,
        month = sep,
       volume = {158},
       number = {3},
          eid = {132},
        pages = {132},
          doi = {10.3847/1538-3881/ab3651},
       adsurl = {https://ui.adsabs.harvard.edu/abs/2019AJ....158..132N}
}

@ARTICLE{Nesvorny2023,
       author = {{Nesvorn{\'y}}, David and {Roig}, Fernando V. and {Vokrouhlick{\'y}}, David and {Bottke}, William F. and {Marchi}, Simone and {Morbidelli}, Alessandro and {Deienno}, Rogerio},
        title = "{Early bombardment of the moon: Connecting the lunar crater record to the terrestrial planet formation}",
      journal = {Icarus},
         year = 2023,
        month = jul,
       volume = {399},
          eid = {115545},
        pages = {115545},
          doi = {10.1016/j.icarus.2023.115545},
       adsurl = {https://ui.adsabs.harvard.edu/abs/2023Icar..39915545N}
}

@ARTICLE{Oro1961,
       author = {{Or{\'o}}, J.},
        title = "{Comets and the Formation of Biochemical Compounds on the Primitive Earth}",
      journal = {Nature},
         year = 1961,
        month = apr,
       volume = {190},
       number = {4774},
        pages = {389-390},
          doi = {10.1038/190389a0},
       adsurl = {https://ui.adsabs.harvard.edu/abs/1961Natur.190..389O}
}

@ARTICLE{Osinski2020,
       author = {{Osinski}, G.~R. and {Cockell}, C.~S. and {Pontefract}, A. and {Sapers}, H.~M.},
        title = "{The Role of Meteorite Impacts in the Origin of Life}",
      journal = {Astrobiology},
         year = 2020,
        month = sep,
       volume = {20},
       number = {9},
        pages = {1121-1149},
          doi = {10.1089/ast.2019.2203},
       adsurl = {https://ui.adsabs.harvard.edu/abs/2020AsBio..20.1121O}
}

@ARTICLE{Patel2015,
       author = {{Patel}, Bhavesh H. and {Percivalle}, Claudia and {Ritson}, Dougal J. and {Duffy}, Colm D. and {Sutherland}, John D.},
        title = "{Common origins of RNA, protein and lipid precursors in a cyanosulfidic protometabolism}",
      journal = {Nature Chemistry},
         year = 2015,
        month = apr,
       volume = {7},
       number = {4},
        pages = {301-307},
          doi = {10.1038/nchem.2202}}

@ARTICLE{Pearce2017,
       author = {{Pearce}, Ben K.~D. and {Pudritz}, Ralph E. and {Semenov}, Dmitry A. and {Henning}, Thomas K.},
        title = "{Origin of the RNA world: The fate of nucleobases in warm little ponds}",
      journal = {Proceedings of the National Academy of Science},
         year = 2017,
        month = oct,
       volume = {114},
       number = {43},
        pages = {11327-11332},
          doi = {10.1073/pnas.1710339114},
       adsurl = {https://ui.adsabs.harvard.edu/abs/2017PNAS..11411327P}
}

@ARTICLE{PierazzoChyba1999,
       author = {{Pierazzo}, E. and {Chyba}, C.~F.},
        title = "{Amino acid survival in large cometary impacts}",
      journal = {Meteoritics \& Planetary Science},
         year = 1999,
        month = nov,
       volume = {34},
       number = {6},
        pages = {909-918},
          doi = {10.1111/j.1945-5100.1999.tb01409.x},
       adsurl = {https://ui.adsabs.harvard.edu/abs/1999M&PS...34..909P}
}

@ARTICLE{Powner2009,
       author = {{Powner}, Matthew W. and {Gerland}, B{\'e}atrice and {Sutherland}, John D.},
        title = "{Synthesis of activated pyrimidine ribonucleotides in prebiotically plausible conditions}",
      journal = {Nature},
         year = 2009,
        month = may,
       volume = {459},
       number = {7244},
        pages = {239-242},
          doi = {10.1038/nature08013},
       adsurl = {https://ui.adsabs.harvard.edu/abs/2009Natur.459..239P}
}

@ARTICLE{RigleyWyatt2022,
       author = {{Rigley}, Jessica K. and {Wyatt}, Mark C.},
        title = "{Comet fragmentation as a source of the zodiacal cloud}",
      journal = {Monthly Notices of the Royal Astronomical Society},
         year = 2022,
        month = feb,
       volume = {510},
       number = {1},
        pages = {834-857},
          doi = {10.1093/mnras/stab3482},
       adsurl = {https://ui.adsabs.harvard.edu/abs/2022MNRAS.510..834R}
}

@ARTICLE{RimmerRugheimer2019,
       author = {{Rimmer}, P.~B. and {Rugheimer}, S.},
        title = "{Hydrogen cyanide in nitrogen-rich atmospheres of rocky exoplanets}",
      journal = {Icarus},
         year = 2019,
        month = sep,
       volume = {329},
        pages = {124-131},
          doi = {10.1016/j.icarus.2019.02.020},
archivePrefix = {arXiv},
       eprint = {1902.08022},
 primaryClass = {astro-ph.EP},
       adsurl = {https://ui.adsabs.harvard.edu/abs/2019Icar..329..124R}
}

@ARTICLE{Rumpf2017,
       author = {{Rumpf}, Clemens M. and {Lewis}, Hugh G. and {Atkinson}, Peter M.},
        title = "{Asteroid impact effects and their immediate hazards for human populations}",
      journal = {Geophysical Research Letters},
         year = 2017,
        month = apr,
       volume = {44},
       number = {8},
        pages = {3433-3440},
          doi = {10.1002/2017GL073191},
       adsurl = {https://ui.adsabs.harvard.edu/abs/2017GeoRL..44.3433R}
}

@ARTICLE{Safrit2021,
       author = {{Safrit}, Taylor K. and {Steckloff}, Jordan K. and {Bosh}, Amanda S. and {Nesvorny}, David and {Walsh}, Kevin and {Brasser}, Ramon and {Minton}, David A.},
        title = "{The Formation of Bilobate Comet Shapes through Sublimative Torques}",
      journal = {Planet Sci. J.},
         year = 2021,
        month = feb,
       volume = {2},
       number = {1},
          eid = {14},
        pages = {14},
          doi = {10.3847/PSJ/abc9c8},
       adsurl = {https://ui.adsabs.harvard.edu/abs/2021PSJ.....2...14S}
}

@ARTICLE{Sasselov2020,
       author = {{Sasselov}, Dimitar D. and {Grotzinger}, John P. and {Sutherland}, John D.},
        title = "{The origin of life as a planetary phenomenon}",
      journal = {Science Advances},
         year = 2020,
        month = feb,
       volume = {6},
       number = {6},
        pages = {eaax3419},
          doi = {10.1126/sciadv.aax3419}
}

@ARTICLE{Sekanina1984,
       author = {{Sekanina}, Z.},
        title = "{Disappearance and disintegration of comets}",
      journal = {Icarus},
         year = 1984,
        month = apr,
       volume = {58},
       number = {1},
        pages = {81-100},
          doi = {10.1016/0019-1035(84)90099-X}
}

@incollection{Shoemaker1962,
      author    = {E. Shoemaker},
      title     = {Physics and Astronomy of the Moon},
      editor    = {A. Kopal},
      booktitle = {Physics and Astronomy of the Moon},
      pages     = {283--571},
      publisher = {Academic Press},
      year      = {1962}
}

@ARTICLE{SpiegelTurner2012,
       author = {{Spiegel}, David S. and {Turner}, Edwin L.},
        title = "{Bayesian analysis of the astrobiological implications of life's early emergence on Earth}",
      journal = {Proceedings of the National Academy of Science},
         year = 2012,
        month = jan,
       volume = {109},
       number = {2},
        pages = {395-400},
          doi = {10.1073/pnas.1111694108},
archivePrefix = {arXiv},
       eprint = {1107.3835},
 primaryClass = {astro-ph.EP},
       adsurl = {https://ui.adsabs.harvard.edu/abs/2012PNAS..109..395S}
}

@ARTICLE{Strom2005,
       author = {{Strom}, Robert G. and {Malhotra}, Renu and {Ito}, Takashi and {Yoshida}, Fumi and {Kring}, David A.},
        title = "{The Origin of Planetary Impactors in the Inner Solar System}",
      journal = {Science},
         year = 2005,
        month = sep,
       volume = {309},
       number = {5742},
        pages = {1847-1850},
          doi = {10.1126/science.1113544},
       adsurl = {https://ui.adsabs.harvard.edu/abs/2005Sci...309.1847S}
}

@ARTICLE{ToddOberg2020,
       author = {{Todd}, Zoe R. and {{\"O}berg}, Karin I.},
        title = "{Cometary Delivery of Hydrogen Cyanide to the Early Earth}",
      journal = {Astrobiology},
         year = 2020,
        month = sep,
       volume = {20},
       number = {9},
        pages = {1109-1120},
          doi = {10.1089/ast.2019.2187}
}

@ARTICLE{Turbet2021,
       author = {{Turbet}, Martin and {Bolmont}, Emeline and {Chaverot}, Guillaume and {Ehrenreich}, David and {Leconte}, J{\'e}r{\'e}my and {Marcq}, Emmanuel},
        title = "{Day-night cloud asymmetry prevents early oceans on Venus but not on Earth}",
      journal = {Nature},
         year = 2021,
        month = oct,
       volume = {598},
       number = {7880},
        pages = {276-280},
          doi = {10.1038/s41586-021-03873-w},
archivePrefix = {arXiv},
       eprint = {2110.08801},
 primaryClass = {astro-ph.EP},
       adsurl = {https://ui.adsabs.harvard.edu/abs/2021Natur.598..276T}
}

@ARTICLE{Way2016,
       author = {{Way}, M.~J. and {Del Genio}, Anthony D. and {Kiang}, Nancy Y. and {Sohl}, Linda E. and {Grinspoon}, David H. and {Aleinov}, Igor and {Kelley}, Maxwell and {Clune}, Thomas},
        title = "{Was Venus the first habitable world of our solar system?}",
      journal = {Geophysical Research Letters},
         year = 2016,
        month = aug,
       volume = {43},
       number = {16},
        pages = {8376-8383},
          doi = {10.1002/2016GL069790},
archivePrefix = {arXiv},
       eprint = {1608.00706},
 primaryClass = {astro-ph.EP},
       adsurl = {https://ui.adsabs.harvard.edu/abs/2016GeoRL..43.8376W}
}

@ARTICLE{Wogan2023,
       author = {{Wogan}, Nicholas F. and {Catling}, David C. and {Zahnle}, Kevin J. and {Lupu}, Roxana},
        title = "{Origin-of-life Molecules in the Atmosphere after Big Impacts on the Early Earth}",
      journal = {Planet. Sci. J.},
         year = 2023,
        month = sep,
       volume = {4},
       number = {9},
          eid = {169},
        pages = {169},
          doi = {10.3847/PSJ/aced83},
       adsurl = {https://ui.adsabs.harvard.edu/abs/2023PSJ.....4..169W}
}

@ARTICLE{Wordsworth2016:EarlyMarsClimate,
       author = {{Wordsworth}, R. D.},
        title = "{The Climate of Early Mars}",
      journal = {Annual Review of Earth and Planetary Sciences},
         year = 2016,
        month = jun,
       volume = {44},
        pages = {381-408},
          doi = {10.1146/annurev-earth-060115-012355}
}

@article{Wyatt2017:ScatteringDesign,
    author      =   {{Wyatt}, M. C. and {Bonsor}, A. and {Jackson}, A. P. and {Marino}, S. and {Shannon}, A.},
    title       =   "{How to design a planetary system for different scattering outcomes: giant impact sweet spot, maximising exocomets, scattered discs}",
    journal     =   {MNRAS},
    volume      =   {464},
    number      =   {3},
    pages       =   {3385--3407},
    year        =   {2017},
    doi         =   {10.1093/mnras/stw2633}
    }

@Article{Xu2018,
    author ="Xu, Jianfeng and Ritson, Dougal J. and Ranjan, Sukrit and Todd, Zoe R. and Sasselov, Dimitar D. and Sutherland, John D.",
    title  ="Photochemical reductive homologation of hydrogen cyanide using sulfite and ferrocyanide",
    journal  ="Chem. Commun.",
    year  ="2018",
    volume  ="54",
    issue  ="44",
    pages  ="5566-5569",
    publisher  ="The Royal Society of Chemistry",
    doi  ="10.1039/C8CC01499J",
    url  ="http://dx.doi.org/10.1039/C8CC01499J",
}

@ARTICLE{Zahnle1986,
       author = {{Zahnle}, Kevin J.},
        title = "{Photochemistry of methane and the formation of hydrocyanic acid (HCN) in the Earth's early atmosphere}",
      journal = {Journal of Geophysical Research: Atmospheres},
         year = 1986,
        month = feb,
       volume = {91},
       number = {D2},
        pages = {2819-2834},
          doi = {10.1029/JD091iD02p02819},
       adsurl = {https://ui.adsabs.harvard.edu/abs/1986JGR....91.2819Z}
}

@ARTICLE{Zahnle2020,
       author = {{Zahnle}, Kevin J. and {Lupu}, Roxana and {Catling}, David C. and {Wogan}, Nick},
        title = "{Creation and Evolution of Impact-generated Reduced Atmospheres of Early Earth}",
      journal = {Planet. Sci. J.},
         year = 2020,
        month = jun,
       volume = {1},
       number = {1},
          eid = {11},
        pages = {11},
          doi = {10.3847/PSJ/ab7e2c},
       adsurl = {https://ui.adsabs.harvard.edu/abs/2020PSJ.....1...11Z}
}

@ARTICLE{Zellner2017,
       author = {{Zellner}, Nicolle E.~B.},
        title = "{Cataclysm No More: New Views on the Timing and Delivery of Lunar Impactors}",
      journal = {Origins of Life and Evolution of the Biosphere},
         year = 2017,
        month = sep,
       volume = {47},
       number = {3},
        pages = {261-280},
          doi = {10.1007/s11084-017-9536-3},
archivePrefix = {arXiv},
       eprint = {1704.06694},
 primaryClass = {astro-ph.EP},
       adsurl = {https://ui.adsabs.harvard.edu/abs/2017OLEB...47..261Z}
}

@ARTICLE{Zellner2020,
       author = {{Zellner}, Nicolle E.~B. and {McCaffrey}, Vanessa P. and {Butler}, Jayden H.~E.},
        title = "{Cometary Glycolaldehyde as a Source of pre-RNA Molecules}",
      journal = {Astrobiology},
         year = 2020,
        month = nov,
       volume = {20},
       number = {11},
        pages = {1377-1388},
          doi = {10.1089/ast.2020.2216}
}

@ARTICLE{Paschek2025,
       author = {{Paschek}, Klaus and {Henning}, Thomas K. and {Molaverdikhani}, Karan and {Miyazaki}, Yoshinori and {Pearce}, Ben K.~D. and {Pudritz}, Ralph E. and {Semenov}, Dmitry A.},
        title = "{Deep Mantle-Atmosphere Coupling and Carbonaceous Bombardment: Options for Biomolecule Formation on an Oxidized Early Earth}",
      journal = {The Astrophysical Journal},
         year = 2025,
        month = may,
       volume = {985},
       number = {1},
          eid = {50},
        pages = {50},
          doi = {10.3847/1538-4357/adc39b},
archivePrefix = {arXiv},
       eprint = {2503.15479},
 primaryClass = {astro-ph.EP},
       adsurl = {https://ui.adsabs.harvard.edu/abs/2025ApJ...985...50P}
}

@ARTICLE{Ferus2020,
       author = {{Ferus}, Martin and {Rimmer}, Paul and {Cassone}, Giuseppe and {Kn\&{\'\i}{\v{z}}ek}, Anton{\'\i}n and {Civi{\v{s}}}, Svatopluk and {{\v{S}}poner}, Judit E. and {Ivanek}, Ond{\v{r}}ej and {{\v{S}}poner}, Ji{\v{r}}{\'\i} and {Saeidfirozeh}, Homa and {Kubel{\'\i}k}, Petr and {Dud{\v{z}}{\'a}k}, Roman and {Petera}, Luk{\'a}{\v{s}} and {Juha}, Libor and {Pastorek}, Adam and {K{\v{r}}ivkov{\'a}}, Anna and {Kr{\^u}s}, Miroslav},
        title = "{One-Pot Hydrogen Cyanide-Based Prebiotic Synthesis of Canonical Nucleobases and Glycine Initiated by High-Velocity Impacts on Early Earth}",
      journal = {Astrobiology},
         year = 2020,
        month = dec,
       volume = {20},
       number = {12},
        pages = {1476-1488},
          doi = {10.1089/ast.2020.2231},
       adsurl = {https://ui.adsabs.harvard.edu/abs/2020AsBio..20.1476F}
}

@article{Knizek2024,
author = {Knížek, Antonín and Petera, Lukáš and Laitl, Vojtěch and Ferus, Martin},
title = {Decomposition of HCN during Experimental Impacts in Dry and Wet Planetary Atmospheres},
journal = {ACS Earth and Space Chemistry},
volume = {8},
number = {6},
pages = {1246-1258},
year = {2024},
doi = {10.1021/acsearthspacechem.4c00064},
URL = {https://doi.org/10.1021/acsearthspacechem.4c00064},
eprint = {https://doi.org/10.1021/acsearthspacechem.4c00064}
}

\section*{Acknowledgements}

The authors thank M. Wyatt, T. Constantinou, and S. White for helpful discussions. 

\section*{Author contributions statement}

Conceptualization, methodology, and visualization: R.J.A., C.H.M, and A.B.
Formal analysis, and software: R.J.A.
Manuscript drafting: R.J.A., C.H.M, A.B., P.B.R., and A.S.P.R.
Project supervision: A.B., and P.B.R.

\section*{Data availability statement}

No new data were generated for this work. Scripts used to generate the figures presented in this work will be made publicly available at \url{https://github.com/richard17a/comet_delivery_hcn}.

\section*{Additional information}

The authors declare that they have no competing interests.

\end{document}